\documentclass[twocolumn]{aastex63}
\usepackage[utf8]{inputenc}
\usepackage[astrosymb]{}
\usepackage{amsmath}
\usepackage{xcolor}
\usepackage{tablefootnote}

\usepackage{xcolor}

\definecolor{dark_red}{RGB}{168, 0, 0}
\definecolor{purple}{RGB}{119, 32, 145}

\received{May 14, 2020}

\submitjournal{AJ}

\shorttitle{PELS-VAE: Generative Modeling of Periodic Variable Stars}
\shortauthors{Martínez-Palomera, Bloom, \& Abrahams}

\graphicspath{{./}{figures/}}

\begin{document}

\title{Deep Generative Modeling of Periodic Variable Stars Using Physical Parameters}

\correspondingauthor{Jorge Martínez-Palomera}
\email{jorgemarpa@berkeley.edu}

\author[0000-0002-7395-4935]{Jorge Mart\'inez-Palomera}
\affiliation{Department of Astronomy, University of California, Berkeley, CA 94720-3411, USA}

\author[0000-0002-7777-216X]{Joshua S.\ Bloom}
\affiliation{Department of Astronomy, University of California, Berkeley, CA 94720-3411, USA}
\affiliation{Lawrence Berkeley National Laboratory, 1 Cyclotron Road, MS 50B-4206, Berkeley, CA 94720, USA}

\author[0000-0002-9879-1183]{Ellianna S. Abrahams}
\affiliation{Department of Astronomy, University of California, Berkeley, CA 94720-3411, USA}
\affiliation{Department of Statistics, University of California, Berkeley, CA 94720-3411, USA}

\begin{abstract}
The ability to generate physically plausible ensembles of variable sources is critical to the optimization of time domain survey cadences and the training of classification models on datasets with few to no labels. Traditional data augmentation techniques expand training sets by reenvisioning observed exemplars, seeking to simulate observations of specific training sources under different (exogenous) conditions. Unlike fully theory-driven models, these approaches do not typically allow principled interpolation nor extrapolation. Moreover, the principal drawback of theory-driven models lies in the prohibitive computational cost of simulating source observables from {\it ab initio} parameters. In this work, we propose a computationally tractable machine learning approach to generate realistic light curves of periodic variables capable of integrating physical parameters and variability classes as inputs. Our deep generative model, inspired by the Transparent Latent Space Generative Adversarial Networks (TL-GANs), uses a Variational Autoencoder (VAE) architecture with Temporal Convolutional Network (TCN) layers, trained using the \hbox{OGLE-III} optical light curves and physical characteristics (e.g., effective temperature and absolute magnitude) from Gaia DR2. A test using the temperature-shape relationship of RR\,Lyrae demonstrates the efficacy of our generative ``Physics-Enhanced Latent Space VAE'' (PELS-VAE) model. Such deep generative models, serving as non-linear non-parametric emulators, present a novel tool for astronomers to create synthetic time series over arbitrary cadences.
\end{abstract}


\section{Introduction} \label{sec:intro}

Robust and in-production automated image-based discovery on streaming survey data has matured significantly, from random-forest based methods \citep{Bloom12,2015AJ....150...82G, 2016ApJ...832..155F,2019PASP..131c8002M}, to deep learning approaches (cf., \citealt{2019A&C....2800284S}). Nonetheless, to extract new knowledge in the time-domain era, the physical nature of the variability must be inferred. Retrospective classification (e.g., after each observing season or after survey completion) has shown great utility for the study of variable stars \citep[e.g.,][]{2005AcA....55...59S, 2013ApJ...763...32D, 2015ApJ...811..113P}. However, the scientific impact for ongoing time-domain surveys such as the Zwiky Transient Factory \citep[ZTF,][]{Bellm_2018}, the Vera Rubin Observatory \citep[VRO-LSST,][]{2019ApJ...873..111I}, and the Wide Field Infrared Survey Telescope \citep[WFIRST,][]{2015arXiv150303757S}, can only be maximized if additional followup resources are appropriately marshalled on scientifically relevant sources. Beyond its utility in broad demographic studies, once a source is classified, inference of the underlying physical state that dictates the observed variability (and any potential differences of that state from others in the same class) is often desirable.

Physical models of transient and variable stars provide, in principle, the most direct path to classification and the inference of the underlying physical state. As generative models---where the relevant initial conditions are fed forward through simulations to obtain the observables---these can be used to solve the inverse problem: the inference of the physical state from the observables. Physical models abound in certain time-domain subfields: e.g., gravitational wave chip signals from binary black hole mergers \citep{2014PhRvD..89d2002K}; the Physics of Eclipsing Binaries \citep[PHOEBE,][]{2016ApJS..227...29P} for binary stars; SNANA \citep{2009PASP..121.1028K} software for supernova analysis; the Modular Open Source Fitter for Transients \citep[MOSFiT,][]{2018ApJS..236....6G} designed for transients interacting with circumstellar material such as  tidal disruption events, kilonovae, Type II supernovae, and Type I superluminous supernova; and PyLIMA \citep{2017AJ....154..203B} for microlensing events. Wrapping physical models within a Bayesian inference framework, e.g., through Markov Chain Monte Carlo (MCMC) modeling, allows one to constrain the parameters of interest with the data.

Physical models, however, present several disadvantages. First, producing observables from {\it ab initio} parameters can be computationally expensive. A generative model which requires even a few seconds of wall-time computation can be prohibitively long when used as part of traditional MCMC inference. This challenge compounds when needing to apply this approach to many sources. Secondly, current models do not include {\it all} physics (intrinsic and extrinsic) and the physical processes that are included are often approximated. As such parameter inference with physical models is inherently imprecise. Last, physical models are often known to describe a subset of the transient and variable stars dynamics. In the absence of a physical model, template fitting based on observed class exemplars may be used. For example, \citet{2010ApJ...708..717S} produced templates of RR\,Lyrae light curves, spanning the range of the observed optical variability. Classification of a new suspected RR\,Lyrae source is then tantamount to a model selection process across the RR\,Lyrae subtypes template bank.

As physical models and templates do not exist generally for the full diversity of the variable sky, classification requires a more data-driven approach. Here, classification is established as a supervised Machine Learning (ML) challenge, where existing data for a set of sources with known classes (``labels'') are used to train an algorithm to predict class membership on new (unlabelled) sources. The efficiency of ML  techniques had been largely demonstrated in providing robust classification of variables sources, either by using feature-based approaches  \citep{2011ApJ...733...10R,Richards12, Pichara13, Pichara16,Nun16, Lochner16, 2018AJ....156..186M} or by directly using the time-series data \citep{2018NatAs...2..151N,2019MNRAS.482.5078A, 2019ApJ...877L..14T, 2020arXiv200308618J}. These retrospective classification efforts\footnote{In contrast, automated streaming machine-learning classification is relatively new: \cite{Muthukrishna+2019, 2019PASP..131j8006C, 2020MNRAS.492.2897Z}; and ALeRCE \url{http://alerce.science/}.} benefited from the use of highly curated training sets.  One principal disadvantage of data-driven (as opposed to physics-driven) classification, however, is the need for a large set of training examples. As new surveys begin, no labeled real data exists with the depth and cadence of the survey\footnote{In the ML context, transfer learning can help address this problem by learning a model using one dataset and predicting in a different domain. For an extensive review see \cite{2019arXiv190304687Z} and \cite{2017ApJ...845..147B} for time-domain astronomy applications.}. Even after a survey has obtained data and sources are labeled, few, if any, of the minority subclasses may be observed and labeled, leading to a large class imbalance that alters the efficacy of classifiers to correctly identify the (often more interesting) minority classes.

To expand the volume of examples in training sets, {\it data augmentation} is often employed \citep{Dieleman15, Cabrera17, 2018AJ....156..186M, 2019AJ....158..257B}. This technique synthesizes new data by generating samples along observational axes believed to be extrinsic to the source itself. Through a series of simple transformations (e.g., rotation, translation, scaling, phase shifting) new  instances are generated. Similarly, for observations with known noise properties, new data can be generated by bootstrap resampling the light curves from the training and/or test datasets (e.g., \citealt{2018NatAs...2..151N}).
Although data augmentation provides a simple and fast path to increase training examples, the methodology expands upon only the known exemplars from the training data. Since the technique exploits a finite set of data, this data-augmentation approach will not generally capture the full continuum of possible behavior within and between classes. That is, from a physical perspective, data augmentation does not afford a principled interpolation nor extrapolation in the way that physics-driven models can naturally accommodate.

Machine learning-based generative modelling, showing recent promise across different domains, provides a more natural framework for improving training set sizes that combine data augmentation techniques with the possibility of interpolation/extrapolation beyond the original training set.
In the ML context, generative models refer to the approach of learning the joint distribution of low-dimensional (latent) random variables that describe the studied phenomena. Deep generative models (DGMs) refer to the use of deep neural network (NN) architectures for the learning and creation process. Multiple variants of DGMs are present in the literature, for a comprehensive review see Chapter 20 of \cite{Goodfellow-et-al-2016}, such as Variational Autoencoder \citep[VAEs,][]{2013arXiv1312.6114K} and Generative Adversarial Networks \citep[GANs,][]{2014arXiv1406.2661G}. Both have shown astonishing results in the image domain, where after training they are able to create realistic new images. Applications of DGMs in astronomy are numerous, and include the works by \cite{2019MNRAS.487L..24T} that exploited both a GAN and a VAE models to map the large-scale gas distribution and temperature of N-body simulations; \cite{2019MNRAS.487.2874I} trained a VAE for anomaly detection in X-ray spectroscopy data; \cite{2019arXiv190906296G} implemented a conditional VAE to speed-up the Bayesian estimation of physical parameters of gravitational wave progenitors; a GAN model for pulsar candidate classification \citep{2019MNRAS.490.5424G}; \cite{2019ComAC...6....1M} used a GAN that generates weak lensing convergence maps; and \cite{2020arXiv200111651Y} trained a VAE model to restore missing data of Cosmic Microwave Background maps.

A major drawback in standard ML generative modelling lies in the limitation of interpolation if unconstrained by physical consideration: generated samples from a learned model may be acceptable visually but are nonetheless unbound to the physics. This shortfall constitutes the starting point for this paper: is it possible to connect the learned latent representation of a generative model with the characteristic/physical attributes of the training data to produce realistic samples that connect to our physical understanding of these sources?

Connecting intrinsic attributes to the latent space has been attempted in the image domain. \citet{2017arXiv170600409L} trained an adversarial encoder-decoder architecture on the CelebA dataset\footnote{CelebFaces Attributes Dataset (CelebA) is a large-scale face attribute dataset with more than 200k celebrity images, each having 40 attribute annotations such as male/female, hair color and length, presence of eyeglasses or hats, nose shapes and smile. \url{http://mmlab.ie.cuhk.edu.hk/projects/CelebA.html}} to disentangle the latent space and the value attributes. The later allows a user of the model to continuously control the parameters of a generated headshot sample. A similar idea was explored by S.\ Guan in his Transparent Latent-space GAN (TL-GAN) \footnote{\url{https://blog.insightdatascience.com/generating-custom-photo-realistic-faces-using-ai-d170b1b59255}}, where he paired a pre-trained GAN model with a pre-trained feature extraction model (both trained with CalebA dataset) to then use a linear regression model to connect the latent space with the predicted features, allowing a smooth exploration of different feature axes (e.g., gender, age, hair type).

Generative models of variable stars capable of reproducing realistic time series can also be an important tool to explore and plan different observation cadences for future time-domain surveys (e.g., VRO-LSST). To optimize cadence strategies, figures of merit must be intercompared with a broad diversity of simulated time-domain sources/events.
The Photometric LSST Astronomical Time-Series Classification Challenge \citep[PLAsTiCC, ][]{2018arXiv181000001T} generated about 3.5$\times 10^6$ light curves, simulated using current physical models and class templates \citep{2019PASP..131i4501K}. Despite the known diversity of galactic variable sources \citep{gaia_collaboration_gaia_2019}, the PLAsTiCC variable stars dataset consisted of just 5 classes (RR Lyrae, Eclipsing Binaries, Miras, microlensing events, and M-dwarf flares). A rational explanation for this is that these are the classes for which reliable physical models and templates exist. The absence of other periodic variables, such as Cepheids and $\delta$\,Scuti, represent the lack of precise physical models that can generate realistic time series. Therefore, this opens a window to the use of state-of-the-art deep learning algorithms to provide fast data-driven non-parametric generative models.

Inspired such challenges faced in massive surveys and the need for expanding the representation across variability classes, here we propose a DGM based in a VAE architecture to simulate new irregularly sampled light curves using physically relevant parameters as input variables. In particular, we train a conditional Variational Autoencoder (cVAE) using OGLE-III light curves for a total of 8 different variability classes. The cVAE model is constructed in two parts: the encoder, that compress the input time series and metadata into a low-dimensional latent vector enclosing the relevant information of each source; and the decoder, which uses the latent code and metadata (the conditional) to reconstruct the original time series. With this design, the model learns the underlying distribution behind the generative process, and therefore is able to create new observations by sampling from the learned distributions. The conditional information provided to the model consists of the variability class, physical parameters, and the time stamp of each observation. We explore the connection between the latent and physical space by means of different regression models. Thus, during evaluation of the model, the user can specify a set of class label, physical parameters, and observation cadence as inputs to the decoder to generate a new realistic light curve.

This paper is structured as follows. Section 2 describes data selection. Section 3 presents the artificial neural network architecture and training procedures. Section 4 discusses the results of the selected generative models. Section 5 presents our conclusions and further prospects.
We include an Appendix that provides a detailed overview of the extensive validation process in our cross-match results.
Alongside this paper, the scripts, trained models, and validated training dataset used for the analysis shown in this work are available online\footnote{\url{https://github.com/jorgemarpa/PELS-VAE}}.

\section{Data} \label{sec:data}

In order to train a DGM that can generate time series of variable sources using physical parameters of the sources as input, we  require a) the light curves of previously classified variable sources and, b) a catalog of relevant physical parameter for these objects, e.g., stellar radius, metallicity, effective temperature. In this section we describe both data sources, as well as data preprocessing.

\subsection{Time Series} \label{sec:lcs}

We  construct our training and testing datasets from The Optical Gravitational Lensing Experiment \citep[OGLE,][]{1992AcA....42..253U} in its third phase \citep[OGLE-III,][]{2008AcA....58...69U}\footnote{\url{http://ogle.astrouw.edu.pl/main/collections.html}}. The \textit{I}-band observed light curves were collected from the Galactic Bulge, Galactic Disk, and the Large and Small Magellanic Clouds fields and describe 8 variability types: Anomalous Cepheids (ACEP), classical Cepheids (CEP), $\delta$\,Scuti (DSCT), Eclipsing Binaries (ECL), Ellipsoidal variables (ELL), Long-Period Variables (LPV), RR\,Lyrae (RRL), and Type II Cepheids (T2CEP). Table \ref{tab:dataset} column (2) summarizes the total number of light curves available in the dataset, as well as the number counts per variability class.

After a visual inspection of the available light curves, we decided to keep only time series with sufficient variability signal. This was assessed by calculating the signal-to-noise ratio between the variability amplitude and the mean photometric uncertainty. We removed light curves with $SNR < 5$, and performed 3 iterations of 3$\sigma$  clipping filter over the magnitude and uncertainty values in order to remove outliers observations. Some of the variability classes have subtypes with multiple pulsation modes and/or are semiregular pulsators, e.g. DSCT-MULTIMODE and LPV-SRV, this results in amorphous shapes in the period-folded (phase) space. We opted to drop these subtypes in order to maintain only variability subtypes that defines a regular shape in their phase-folded light curve. The subtypes that most impacts the volume of our dataset are LPVs semi regular variables (SRV), OGLE small amplitude red giants (OSARG), carbon rich (C), and oxygen rich (O) variables.

\begin{figure*}[htb!]
\plotone{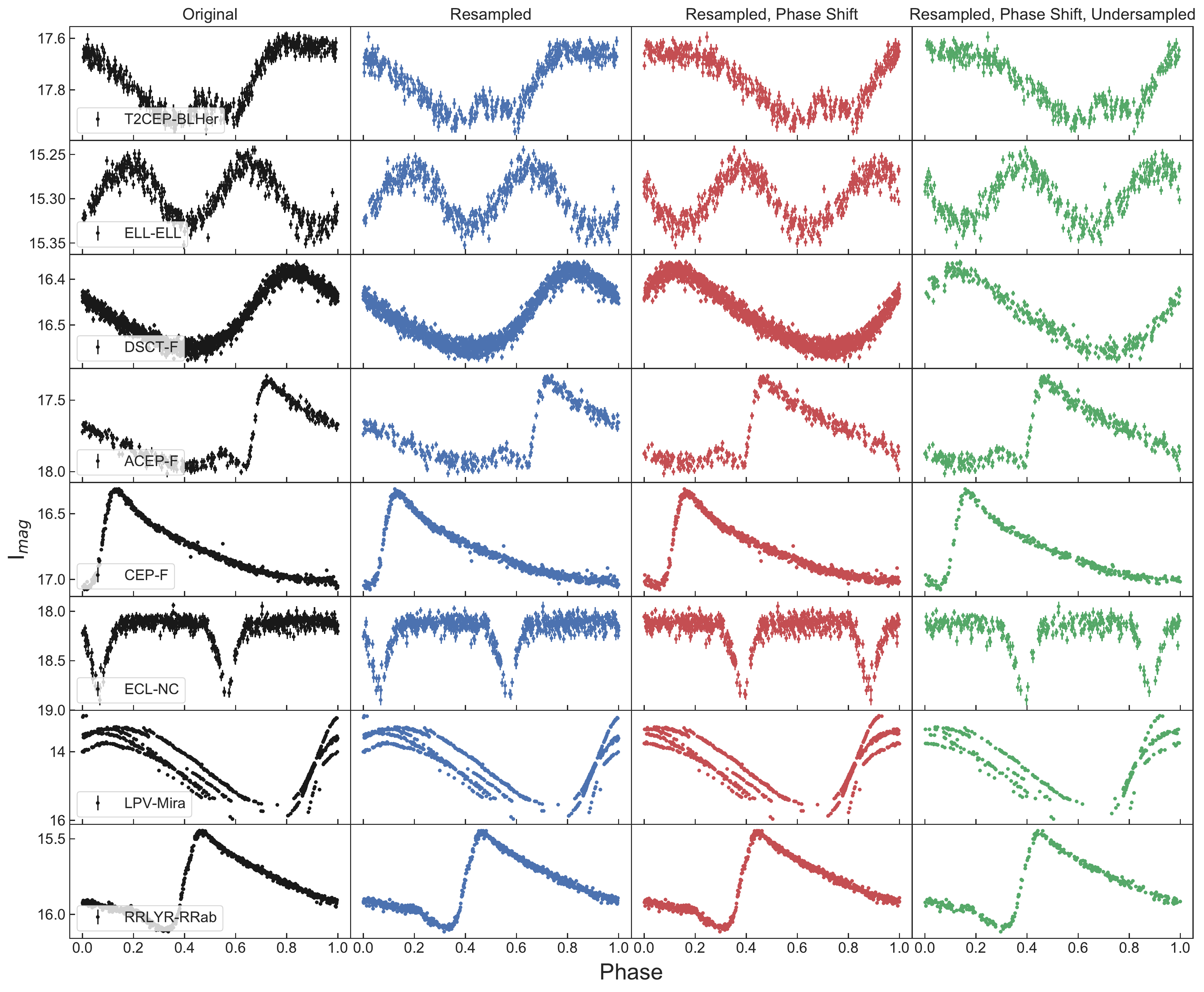}
\caption{Light curve examples from OGLE-III survey. Eight different variability classes are shown in each row. In column 1 the original phase folded light curves are shown; column 2 shows the resampled light curves, column 3 shows the phase-shifted light curve, and column 4 depicts the undersampling to $t_{\rm len}=300$ observations.
\label{fig:aug_ex}}
\end{figure*}

In order to train our proposed NN (see Section \ref{sec:mod_arch}), we required all light curves to have the same number of observations. After analysing the distributions of light-curve length of OGLE-III data, we decided to set the sequence length to $t_{\rm len}=300$ observations. We phase-folded every time series using the periods reported by OGLE-III studies and then we randomly undersample to $t_{\rm len}$ data points per light curve.
The total number counts after the preprocessing step is presented in Table \ref{tab:dataset} columns (3), light-curve examples are shown in column (1) of Figure \ref{fig:aug_ex}.

\begin{deluxetable*}{lrrrrrrrr}
\tablecaption{OGLE-III Light Curves and GAIA DR2 stellar parameters. \label{tab:dataset}}
\tablewidth{0pt}
\tablehead{
\colhead{Variable} & \multicolumn{4}{@{}c}{Total Light Curves} & \multicolumn{2}{@{}c}{$T_{\rm eff}$} & \multicolumn{2}{@{}c}{R $\&$ L} \\
\colhead{} & \colhead{\textit{Original}\tablenotemark{a}} & \colhead{\textit{Clean}} & \colhead{\textit{Validated}} & \colhead{\textit{Augmented}} & \colhead{\textit{Validated}}  & \colhead{\textit{Augmented}} & \colhead{\textit{Validated}}  & \colhead{\textit{Augmented}}
}
\colnumbers
\startdata
 ACEP    &       83 &     72 &    71  &  5,000 &     1 ( 1.4$\%$) &     70 ( 1.4$\%$) &     -    (0$\%$) &     -    (0$\%$) \\
 CEP     &    8,052 &  7,265 & 7,121  & 10,045 & 4,931 (69.3$\%$) &  6,934 (69.0$\%$) &     5  (0.1$\%$) &     6  (0.1$\%$) \\
 DSCT    &    2,596 &     44 &    42  &  5,090 &    32 (76.2$\%$) &  3,840 (75.4$\%$) &    10 (23.8$\%$) & 1,166 (22.9$\%$) \\
 ECL     &  419,868 & 10,002 & 9,505  & 10,000 & 8,581 (90.3$\%$) &  9,035 (90.4$\%$) & 1,495 (15.7$\%$) & 1,556 (15.6$\%$) \\
 ELL     &   25,217 &  2,328 & 2,269  & 10,365 & 1,908 (84.1$\%$) &  8,720 (84.1$\%$) &   135  (5.9$\%$) &   603  (5.8$\%$) \\
 LPV     &  343,596 &  4,460 & 4,349  & 10,044 & 3,730 (85.8$\%$) &  8,638 (86.0$\%$) &     6  (0.1$\%$) &    15  (0.1$\%$) \\
 RRLYR   &   44,031 & 10,028 & 9,322  & 10,169 & 2,814 (30.2$\%$) &  3,062 (30.1$\%$) &    31  (0.3$\%$) &    31  (0.3$\%$) \\
 T2CEP   &      599 &    450 &   436  &  5,047 &   322 (73.8$\%$) &  3,746 (74.2$\%$) &     3  (0.7$\%$) &    32  (0.6$\%$) \\
\hline
 Total   &  844,042 & 58,049 & 33,114 & 65,760 & 22,319(67.4$\%$) & 44,045 (70.0$\%$) & 1,685 (5.1$\%$) &  3,409 (5.2$\%$) \\
\enddata
\tablenotetext{a}{Includes all variability subtypes.}
\tablecomments{The original number of light curves available from OGLE-III database are shown in column (2). Columns (3) and (4) give the number of sources after the cleaning and cross-match validation \ref{apen:x-match} for details), respectively. Column (5) shows the number of sources after augmentation. Column (6) and (7) show the number of sources with $T_{\rm eff}$ values (percentages) for the validated cross-matches and augmented dataset, respectively. Similarly, columns (8) and (9) show the stellar radius and luminosity.}
\end{deluxetable*}

\subsection{Data Augmentation} \label{sec:augment}

As seen in Table \ref{tab:dataset} (column 3), the variability classes distribution reflects of a clear imbalance, with a conspicuous lack of sources in the DECT, ACEP, and T2CEP classes. In order to compensate for this we artificially augmented the dataset. For a given phase-folded non-trimmed light curve we resample the photometric measurements following a Gaussian distribution with mean and variance corresponding to the photometric magnitude value and its associated uncertainty, respectively. Then we applied a phase shift sampled from a [0,1] uniform distribution. Finally, we randomly undersample the new light curve to $t_{\rm len}$ observations.
We performed these steps for a random selection of sources in each variability class to reach a uniform count of data per class. The total number of time series per class after data augmentation is shown in column (5) of Table \ref{tab:dataset}. For classes with more than two thousand examples, we augmented the number counts up to $\sim$10,000, for the rest to $\sim$5,000. In the case of ECL, where the initial number of sources is noticeably larger that for other types, we reduced the dataset to $\sim$10,000 examples prioritizing sources with physical parameters (see Section \ref{sec:phyparams}).
Figure \ref{fig:aug_ex} shows light curves for different variability classes, as well as the result of the three steps followed in our augmentation procedure. This demonstrates that the important characteristics of each light curve such as shape, amplitude, and photometric statistics are reasonably preserved.

\subsection{Physical Parameters} \label{sec:phyparams}

\begin{figure}[htb!]
\plotone{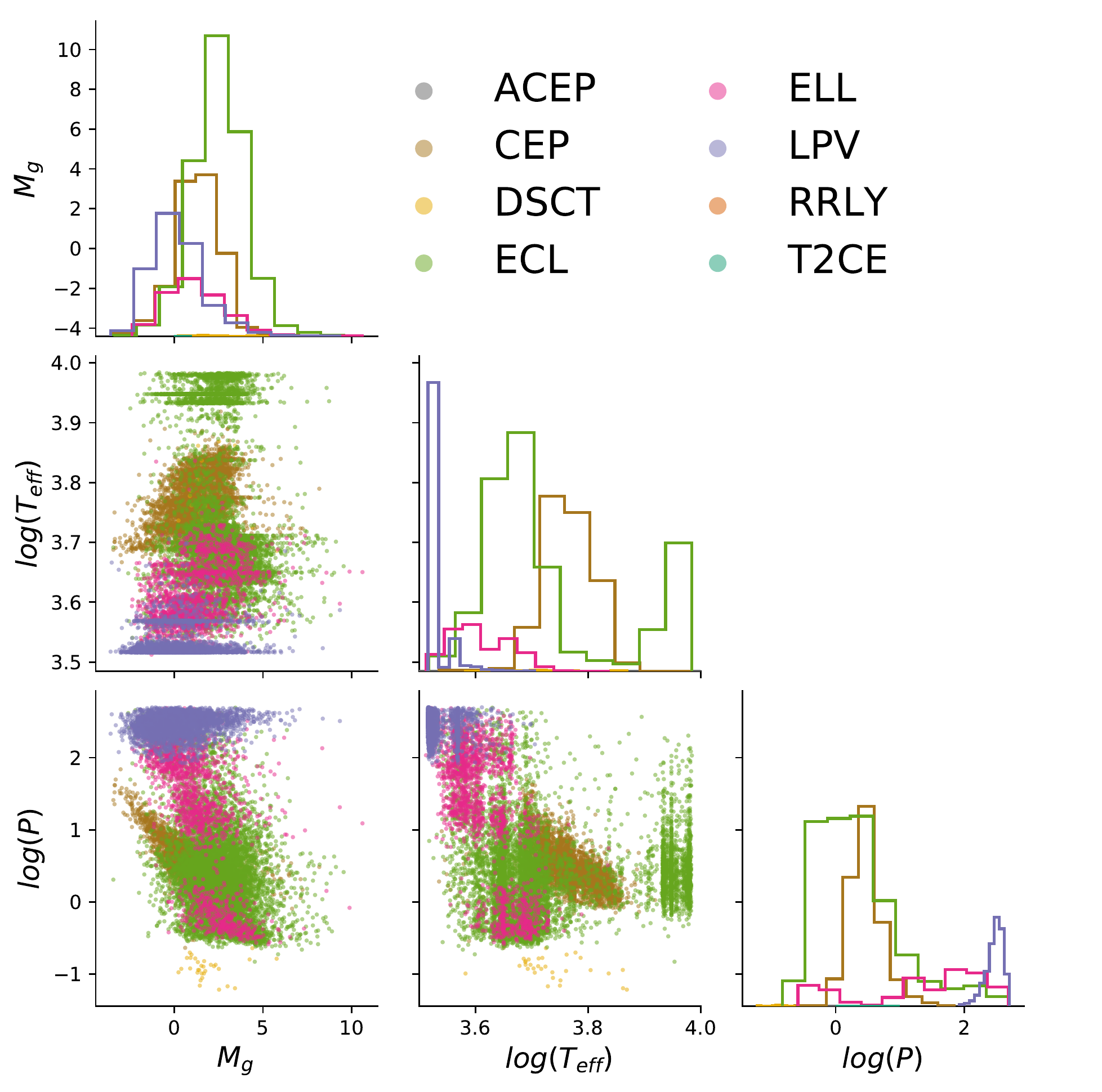}
\caption{Joint distributions of physical parameters period $P$, effective temperature $T_{\rm{eff}}$, and absolute g-band magnitude $M_g$ used during training of the generative model (see Section \ref{subsec:emb_pp}) color coded by variability class.
\label{fig:pp_joint}}
\end{figure}

To incorporate stellar parameters in our model we cross-matched the OGLE-III variability catalog with the Gaia Data Release 2 \citep[GAIA DR2,][]{2016A&A...595A...1G,2018A&A...616A...1G} using a $2 \arcsec$ radius search. We followed a rigorous set of steps in order to validate the cross-matched sources (more details in Appendix \ref{apen:x-match}), which include compensating for proper motion, comparison of GAIA and OGLE variability classification, and positioning of variables in the color-magnitude diagram. This provides stellar parameters for a fraction of the light-curve dataset, such as: effective temperature $T_{\rm eff}$, stellar radius $R$, and luminosity $L$. Table \ref{tab:dataset} shows the number of sources with stellar parameters of all eight variability classes. We would like to highlight that a sizeable fraction (38$\%$) of our dataset is associated with measures of effective temperatures, whereas few measurements of stellar radii and luminosities from Gaia are made available. The Gaia pipeline only provides $R$ and $L$ for less than half of the sources with temperature estimates due to post-processing filtering \citep{2018A&A...616A...8A}. Due to the lack of sources with stellar luminosity and radius, we choose to exclude these physical parameters in our training process. The Gaia catalog also provides color values based in the blue and red passband ($\rm{G}_{\rm{BP}} - \rm{G}_{\rm{RP}}$) and parallaxes \citep{2019A&A...623A.110G}, absolute \textit{g}-band magnitude M$_{\rm{G}}$ were calculated using distances derived by \cite{2018AJ....156...58B}. Figure \ref{fig:pp_joint} shows the joint distribution of the three physical parameters (period, M$_{\rm{G}}$, and $T_{\rm eff}$) used during model training, see Section \ref{subsec:emb_pp} for details.

\section{Neural Network Model and Training} \label{sec:mod_arch}

\subsection{Network Architecture}

We used the Variational Autoencoder \citep[VAE,][]{2013arXiv1312.6114K} architecture as our deep generative model of choice. A VAE provides a probabilistic approach for calculating a compressed representation of a set of observations.
A VAE is described by two components. First, an encoder stage transforms training data into a low-dimensional representation in the so-called latent space. Then, a decoder processes the latent representation and expands it in order to reconstruct the original data.
In a VAE, the encoder output describes a probability distribution for each latent dimension, instead of a deterministic representation as the case of classic autoencoders \citep{2006Sci...313..504H}. The dimensionality of the latent space is a hyperparameter of the model to tune. This probability distribution is assumed to be normally distributed and the encoder predicts its mean and variance values. Later, a latent vector $z$ is sampled from the learned distribution and fed into the decoder using the reparameterization trick:
\begin{equation}
\label{eq:reparam}
z = \mu + \sigma \odot \epsilon
\end{equation}
where $\mu$ and $\sigma$ describe the probability distribution returned by the encoder, and $\epsilon$ is sampled from a unit Gaussian distribution. This allows performing backpropagation during the training phase.

\begin{figure*}[ht!]
\plotone{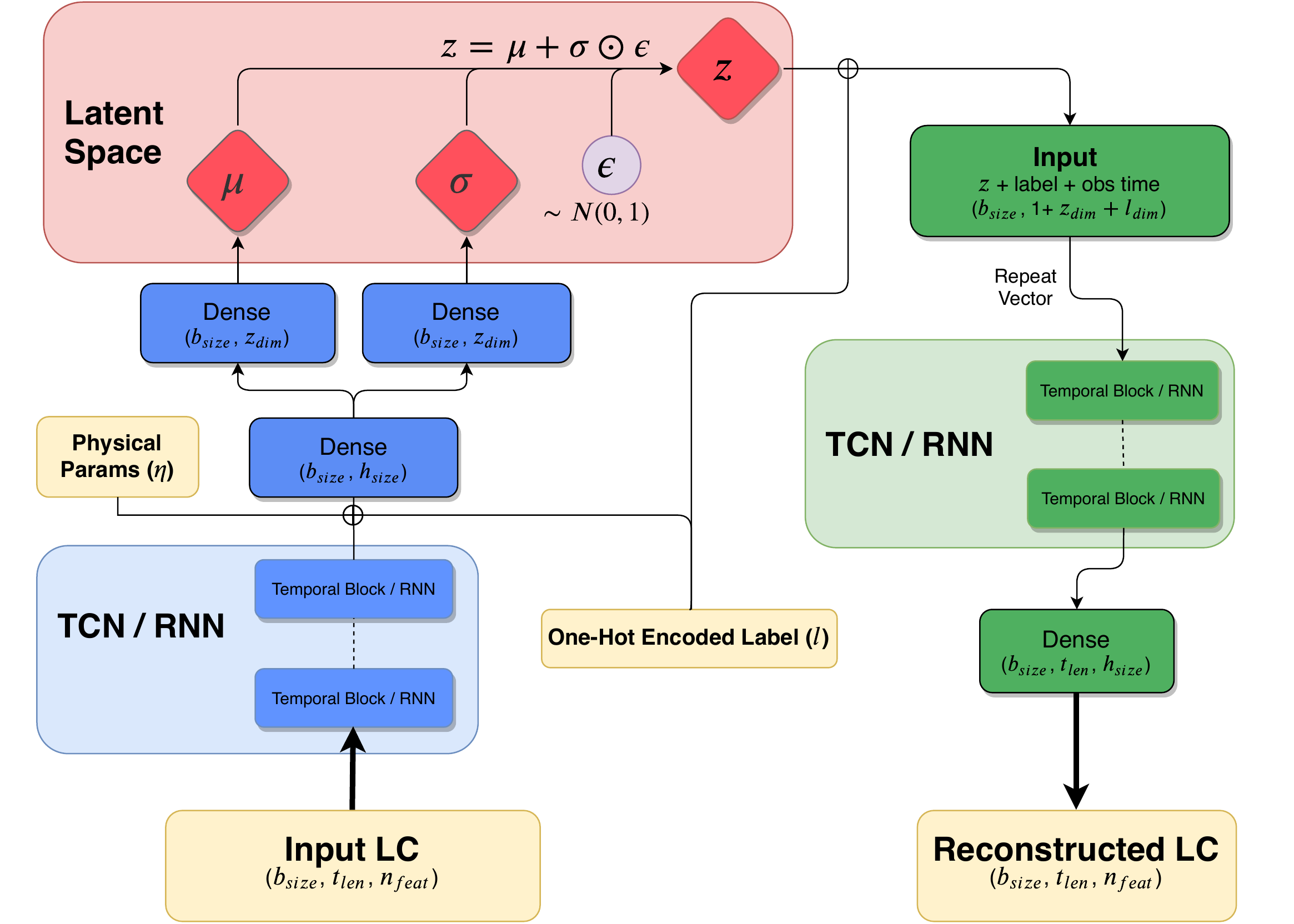}
\caption{Neural network architecture of a conditional Variational Autoencoder (cVAE). The left hand side of the diagram represents the encoder (blue boxes), while the right hand-side represents the decoder (green boxes). Red boxes shows the latent space. Each sides, encoder and decoder, use a sequential architecture (TCN or RNN) and a fully connected dense layer to map the outputs. Yellow boxes represent input data, such as, light curves (LC), variability labels ($l$), and the corresponding physical parameters ($\eta$).
\label{fig:vae_net}}
\end{figure*}

Our encoder-decoder architecture consists of two types of layers for each module, a temporal layer processing the sequential nature of the data, and fully connected layers for outputs. Figure \ref{fig:vae_net} shows an overview of the VAE network architecture. The temporal component can be implemented as either a Temporal Convolutional Network \citep[TCN,][]{2018arXiv180301271B}, or a Recurrent Neural Network (RNN).

TCNs refer to a family of 1-D convolutional architectures designed for efficiently handling sequential data. The main features of TCNs are 1) causal convolutions, meaning that there is no information leakage from future to past; 2) dilated convolutions, the equivalent of adding a step between every adjacent filter to allow for a large receptive field \citep{2015arXiv151107122Y} and an extensive lookback time; and 3) residual connections, where the output of each residual block is constructed by adding the input data and the transformed data layers \citep[see Figure 2 in][]{He_2016_CVPR}. TCN networks are described by the following hyperparameters: the convolution kernel size ($k_{\rm{size}}$), the number of hidden units ($h_{\rm{size}}$) in the convolutional layers, the dilation of convolution ($d$), and the number of temporal blocks ($n_{\rm{blocks}}$).

Alternatively, RNNs are recursive architectures that combine operations per-cell (time step) in order to calculate a cell-state and output. Such cell-states are carried into the next cell (time step) and contain the relevant historic information learned. RNNs suffer from several well-known issues such as short-term memory and vanishing gradient problem during training \citep{279181, 2012arXiv1211.5063P}. There are variants of RNN architectures designed to prevent such limitations. The most widely used are Long-Short Term Memory \cite[LSTM,][]{lstm} and the Gated Recurrent Units \cite[GRUs,][]{2014arXiv1406.1078C} that proved to be performant to prevent the gradient vanishing and explosion typically noted in traditional RNNs, by including an internal mechanism called gates to regulate the flow of information. The network size is controlled by the number of hidden units per cell ($h_{\rm{size}}$) and the number of stacked RNNs ($n_{\rm layers}$), where the output of each cell is fed into the next cell in the same RNN layer but also to the corresponding cell in the next RNN layer.

After the sequential layer(s) in the encoder we include stacks of fully connected layers followed by a ReLU\footnote{A Rectified Linear Unit (ReLU) function is defined as \hbox{$f(x) = \max(0,x)$}.} activation and dropout layer\footnote{Dropout is a regularization technique and refers to the process of randomly deactivating neurons during training in order to avoid overfitting. The number of dropped neurons per layer is defined by a probability which is an hyperparameter of the model.}. The networks then connects to two independent fully connected layers, one to predict the mean and the other the log-variance of the $n$-dimensional Gaussian distributions of the latent space.

For the decoder, a sequential network (TCN or RNN) receives as input a repeated vector of the latent code and meta-data reshaped according to the number of time steps $t_{\rm len}$. Each time step is tagged with the corresponding observed difference in time $\Delta t_i = t_i - t_{i-1}$. This sequential network uses the same architecture and hyperparameters as the encoder. After that, a fully connected layer followed by a sigmoid\footnote{A sigmoid function is defined as $f(x) = ({1 + e^{-x}})^{-1}$ and constrains output values to [0,1] range.} activation function returns the reconstructed scaled light curve.

The data flow thorough the VAE network (see Figure \ref{fig:vae_net} for reference, where the arrows represent the flow of data) follows as: scaled light curves are first fed into the encoder leading to the extraction of representative features. Then the last time-step state is concatenated with a one-hot encoding of the label value and the physical parameters. Next, the stack of fully connected layers is branched into the two dense layers that predict the mean and log-variance of the latent space distributions. Later, a new latent vector is sampled and concatenated with the observations times of the light curve (as in \citealt{2018NatAs...2..151N}), and the encoded labels; this vector is repeated $t_{\rm len}$ times and presented to the sequential component of the decoder; finally a fully connected layer processes the sequential output and returns the magnitude and error of the reconstructed light curve.
Our VAE model accepts non-uniformly sampled time series and is time conditioned, therefore only reconstructs the photometric measurements. Moreover, due to the inclusion of side information into the network, the latent variables not only encodes the relevant features extracted from the light curves, but also embed the provided metadata, which enforces a correlation between the latent space and the physical parameters that can be exploited after training. We call this architecture the ``Physics-Enhanced Latent Space VAE'' (PELS-VAE) model.

\subsection{Training} \label{sec:training}

Let $x$ be the observed training datapoints, $z$ the latent vector, $q_{\theta}(z|x)$ the encoder network with $\theta$ model parameters, and $p_{\phi}(x|z)$ the decoder network with $\phi$ model parameters, then the classical VAE objective function is:
\begin{equation}
\label{eq:vvae_loss}
\mathcal{L} = \mathbb{E}_{z \sim q_{\theta}(z|x)} [\log p_{\phi}(x|z)] - D_{KL}(q_{\theta}(z|x)||p(z))
\end{equation}
where the first term is the reconstruction likelihood of the decoder network given a latent vector, the expectation value is taken with respect the encoder's distribution over the representations.
The second term is a regularization, the Kullback-Leibler divergence \citep[KL,][]{kullback1951} between the learned latent distribution and its prior, which is assumed to be the unit Gaussian $p(z) \equiv \mathcal{N}(0,\mathbf{I})$, with $\mathbf{I}$ the identify matrix with dimension corresponding to the size of the latent space.
This loss function equally treats the reconstruction error and the similarity of the latent representation with a unit Gaussian. The latter intends to capture the underlying data generative factors, enforce that similar datapoints have a similar latent representation, and aims for a disentangled representation, meaning that each single latent directions controls a single aspect of the generative factor.
One way to enforce disentanglement in the latent space would be done by introducing an additional hyperparameter ($\beta$) that weights the importance of the second term in equation \ref{eq:vvae_loss} as follows:
\begin{equation}
\label{eq:bvae_loss}
\mathcal{L} = \mathbb{E}_{q(z|x)} [\log p(x|z)] - \beta D_{KL}(q(z|x)||p(z))
\end{equation}
Introduced by \cite{2018arXiv180403599B}, the hyperparameter $\beta$ plays a role in disentangling the latent representation. Higher values of $\beta$ enforce orthogonality between latent directions due to the assumption of a diagonal covariance matrix in its prior distribution. With $\beta = 0$, the traditional autoencoder loss is recovered.
We used a slightly modified version of the empirical expression for Equation \ref{eq:bvae_loss} when $p(z) \equiv \mathcal{N}(0,\mathbf{I})$:
\begin{eqnarray}
\label{eq:empi_loss}
  \mathcal{L} = \frac{1}{t_{\rm len} N} \sum_{i=0}^{N}\sum_{j=0}^{t_{\rm len}}\left(\frac{x^{j}_{i} - \hat{x}^{j}_{i}}{\sigma^j_{i}}\right)^2 \nonumber \\
- \beta \sum_{i=0}^{N} (\sigma_i^2 + \mu_i^2 - \log(\sigma_i) - 1) \nonumber \\
+ D_{KL}(\hat{\sigma}^j_i||\sigma^j_i)
\end{eqnarray}
where $N$ is the total number of light curves, $x^j_i$, $\hat{x}^j_i$, $\sigma^j_i$, and $\hat{\sigma}^j_i$ are the $j$th measurements, reconstruction values, measurement error, and reconstructed errors of the $i$th light curve, respectively, $\mu_i$ and $\sigma_i$ refer to the mean and dispersion the latent distributions for the $i$th light curve. The first term correspond to the weighted mean squared error that depicts the reconstruction error, and the second term refers to the KL divergence.
We added a third regularization term that enforce the proper reconstruction of predicted measurement errors by calculating the KL divergence between the true and predicted values. This last term regularizes that the probability distribution of reconstructed errors $\hat{\sigma}^j_i$ follows the true distribution of $\sigma^j_i$.

We partition our dataset into three subsamples, the training (60$\%$), validation (20$\%$), and test (20$\%$) sets. We followed a stratified split strategy to ensure that class proportions are preserved for each partition. The test set only contains real sources that were not used during data augmentation.
To search for the hyperparameters set of the best-performing model, we run a hyperparameter sweep and optimization using the Weight \& Biases\footnote{\url{https://www.wandb.com/}} framework. We used a Bayesian Optimization search strategy provided by such framework that employs a Gaussian process to model the hyperparameter function and then chooses parameters that improves the probability of minimizing a specific metric, which in our case was the loss function for the validation set. The hyperparameter search covered different combinations and are summarized in Table \ref{tab:arch_params}.

\begin{deluxetable*}{cll}
\tablecaption{Neural Network hyperparameters notation and grid search
\label{tab:arch_params}}
\tablewidth{0pt}
\tablehead{
\colhead{Parameter} & \colhead{Description} &\colhead{Grid search$^{*}$}
}
\startdata
 $b_{\rm{size}}$ & Batch size & [32, 64, \textbf{128}] \\
 $lr$ & Learning rate & $\sim U(0.00005, 0.1)$, \textbf{0.001}\\
 $lr_{\rm{sch}}$ & Learning rate scheduler & [None, Exponential, \textbf{Cosine}, Plateau] \\
 $\beta$ & KL divergence weight & $\sim U(0,1)$, \textbf{0.75} \\
 $z_{\rm dim}$ & Latent space dimension & [\textbf{4}, 6, 8, 12] \\
 $p_{\rm drop}$ & Dropout probability & $\sim U(0, 0.5)$, \textbf{0.2} \\
 \hline
 $n_{\rm blocks}$ & Number of temporal blocks in TCN & [5, \textbf{7}, 9] \\
 $k_{\rm size}$ & TCN kernel size & [3, 5, \textbf{7}, 9] \\
 $d$ & Dilation in TCN & 2 \\
 seq\_arch & Sequential architecture & \textbf{TCN}, GRU, LSTM \\
 $h_{\rm size}$ & Number of hidden units in TCN/RNN & [16, 32, \textbf{48}, 64] \\
 $n_{\rm layers}$ & Number of RNN layers & [1, \textbf{2}, 3] \\
\enddata
\tablecomments{$^{*}$Hyperparameters grid. In bold, are highlighted the values associated to the best-performing model.}
\end{deluxetable*}

After the sweep search, we found the set of optimized hyperparameters highlighted in Table \ref{tab:arch_params}. We treated the latent space dimension as a model's hyperparameter and optimized for best loss performance, which makes it model and dataset dependent. The best latent dimensions are 4 and 6, with insignificant differences in their loss performance, but with higher correlation coefficients between embeddings for the later. Therefore, a 4-dimension latent space is sufficient to encapsulate the necessary information to then fully reconstruct the original time series, while still keeping a low-dimensional space that can be correlated to a low-dimensional physical parameter space {\it a posteriori}. We did not find a significant difference between the best configuration of GRU and TCN in terms of reconstructed light curves and latent space properties, but found a reduced convergence time in training for TCNs, which were at least 3 times faster than GRUs, even tough the TCN network capacity was 6 times larger. This is consistent with the recent findings in \citet{2020arXiv200308618J}.

\begin{figure}[ht!]
\plotone{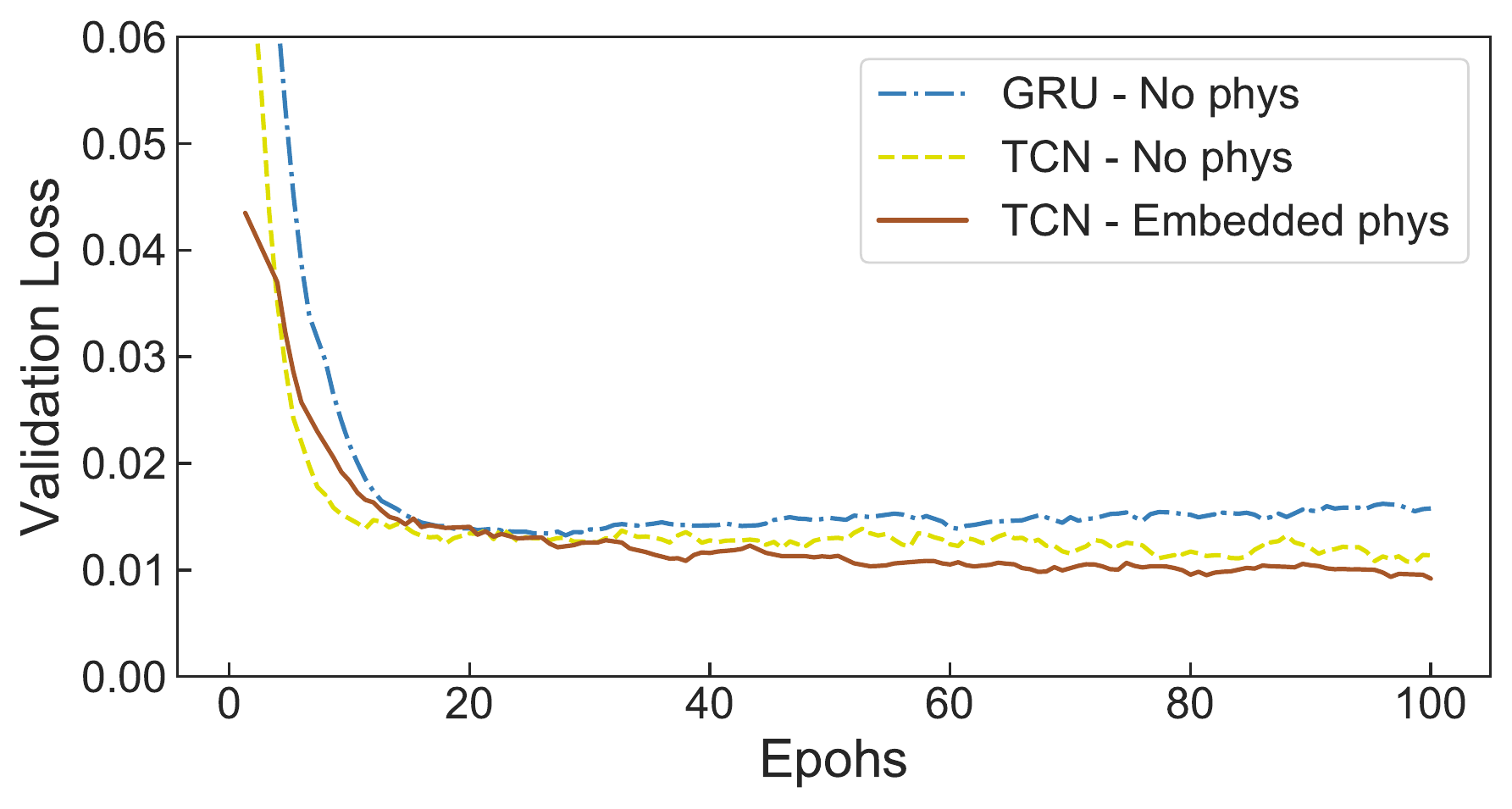}
\caption{Validation loss during training epochs for VAE models without physical parameters and GRU (blue), TCN (blue) architectures, while the red line shows a model with TCN layers that include physical parameters during training.
\label{fig:val_loss}}
\end{figure}

We used the ADAM optimizer \citep{2014arXiv1412.6980K} during training over 100 epochs. Training and testing loss values are shown in Figure \ref{fig:val_loss}, where convergence is shown to be achieved.
Our models were implemented using \texttt{Pytorch\,1.3} \citep{paszke2017automatic}.

\section{Results} \label{sec:resutls}

\subsection{Light-curve Reconstruction} \label{subsec:lc_recon}

\begin{figure*}[ht!]
\plotone{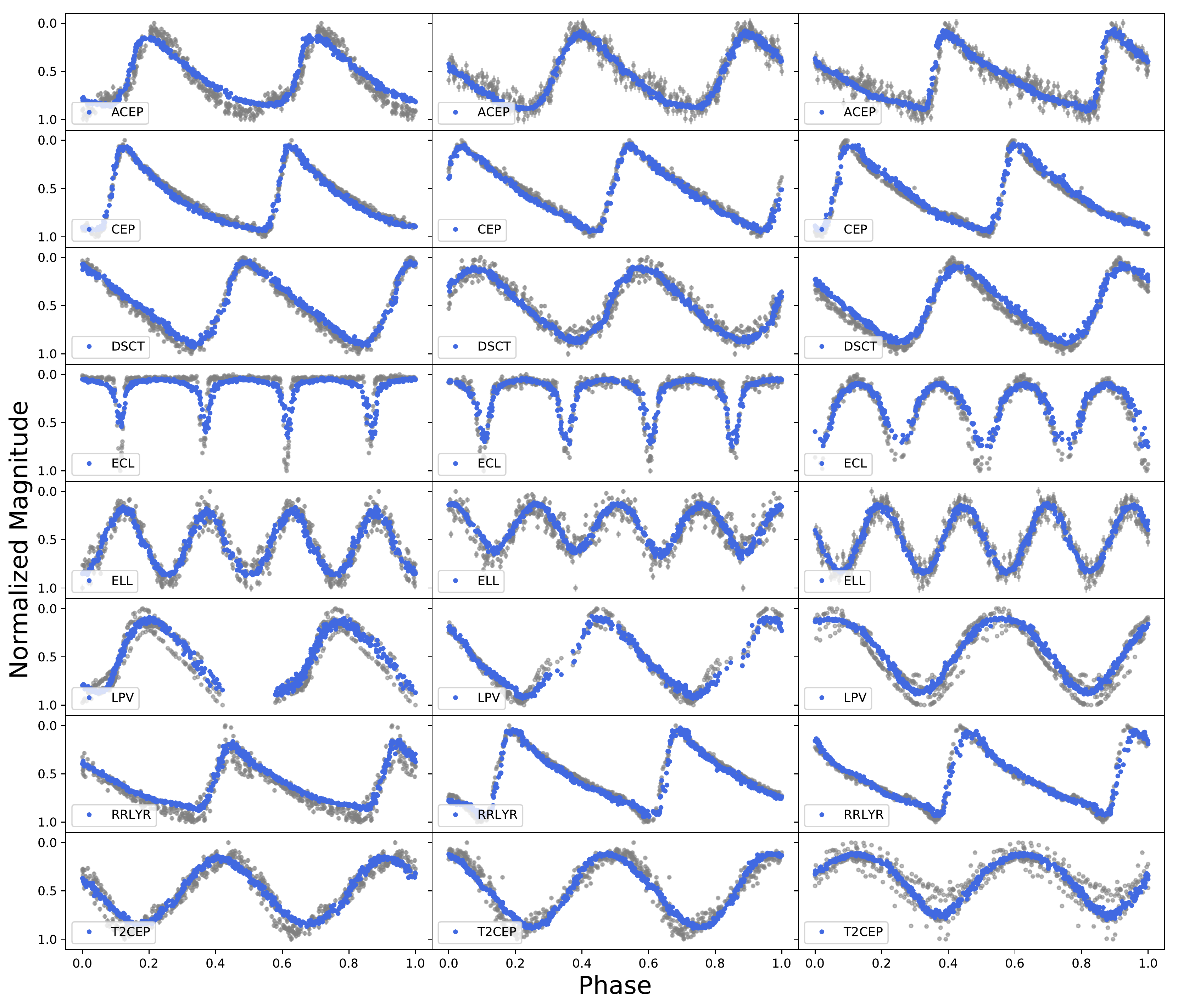}
\caption{Displays of reconstructed phase-folded light-curves obtained at the decoder level by the best-performing model trained only with time series (cVAE). Three examples per variability class are shown. Gray markers denote the observed photometric OGLE-III I-band light curve, while in blue are the decoder reconstructions.
\label{fig:lc_rec_no}}
\end{figure*}

First we explore the basic form of our cVAE model, trained with light curve data and conditioned only with variability labels, excluding physical parameters, in order to explore the capabilities of the model to capture the necessary information to reconstruct light curves of periodic variable sources.
Figure \ref{fig:lc_rec_no} shows reconstructed light curves from the selected best model, with three examples for every variability class. The overall shape and small details characteristic to each variability class are recovered by the generative model. Due to the variational nature of the model, stochasticity introduced when sampling from the latent variables, the reconstructed light curves are not completely equal to their original counterparts. As expected, the model is optimize to learn a smooth latent space that facilitates the generative process rather than the reconstruction.

\begin{figure*}[ht!]
\plotone{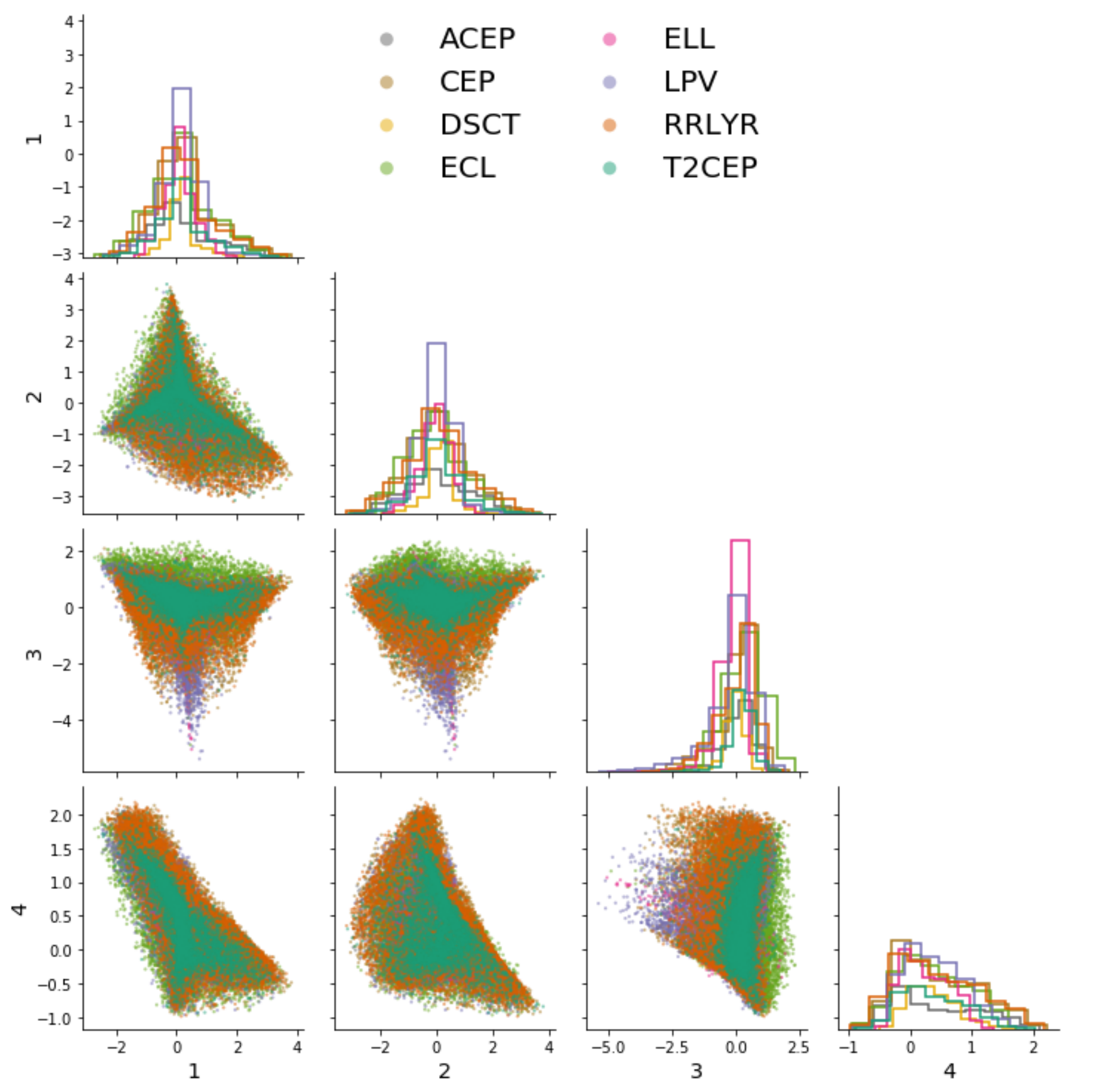}
\caption{Joint distributions of all $\mu$ (encoded features) values for the 4 dimensions of the latent space obtained by the best-performing model trained only with time series data (cVAE). Color coding correspond to the 8 variability types.
\label{fig:pair_latent_no}}
\end{figure*}

Figure \ref{fig:pair_latent_no} shows the joint distributions of all four latent dimensions, particularly the predicted mean values ($\mu$) that describes the Gaussian distribution of the latent space. Due to the regularization term added to the objective function, the KL divergence term in eq.\,\ref{eq:vvae_loss}, the learned latent space resembles a normal distribution in each dimension. Clustering of different variability classes is not strong, due to this regularization, which drives towards a smooth and dense latent space. The later is particularly useful when generating new instances, especially when interpolating between different loci of the latent space that were not explored during training.

\subsection{Embedding Physical Parameters} \label{subsec:emb_pp}

\begin{figure*}[ht!]
\plotone{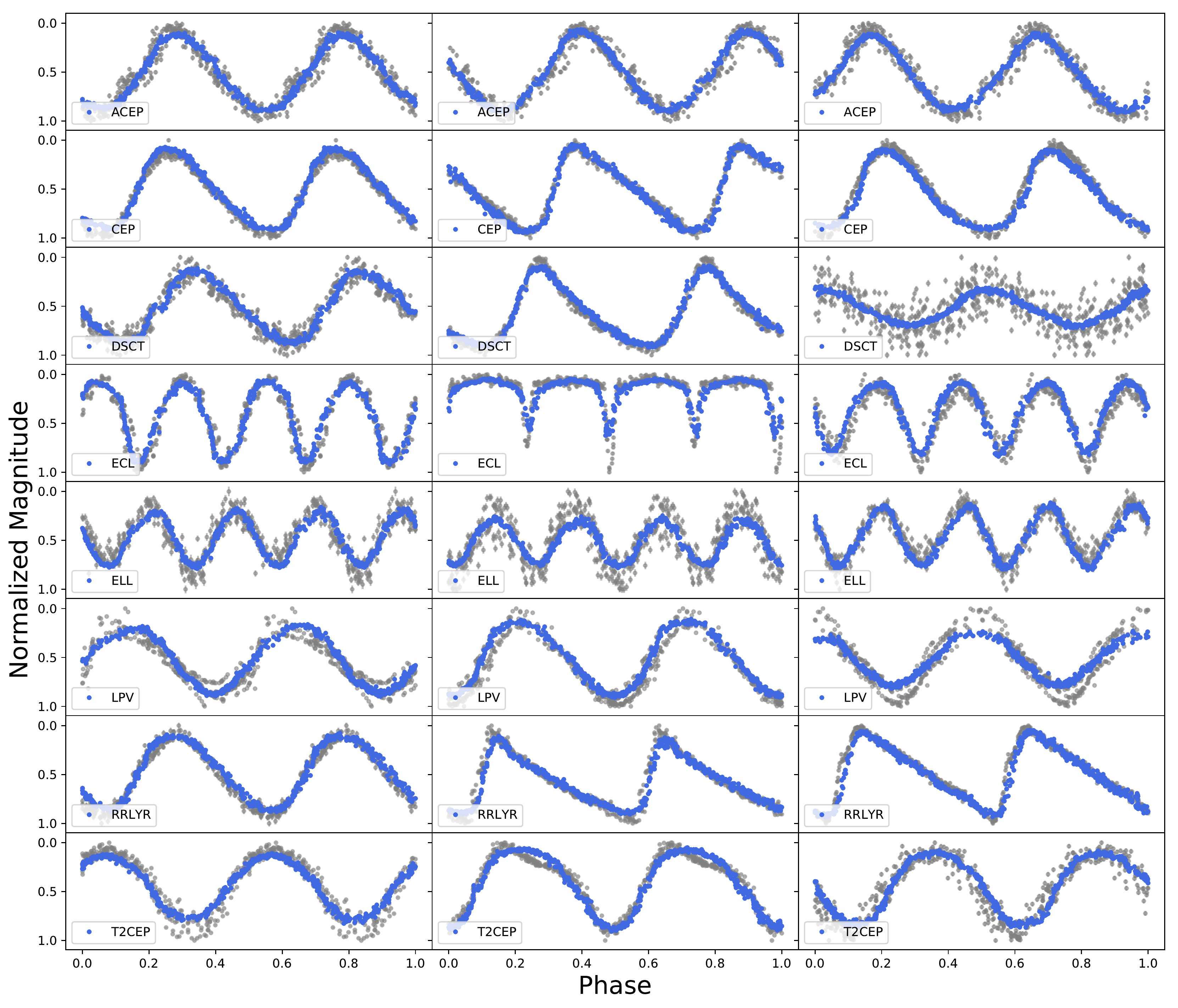}
\caption{Reconstructed light-curves obtained at the decoder level by the best-performing network (cVAE-P) exploiting physical parameters as auxiliary inputs. Displays of 3 objects per variability class are shown. Gray markers denote of the observed photometric OGLE-III I-band light curve, and in blue the decoder reconstructions.
\label{fig:lc_rec_yes}}
\end{figure*}

Using the same network architecture described in Figure \ref{fig:vae_net}, we trained a second model (cVAE-P) that includes three physical parameters: effective temperature, absolute magnitude, and period. Figure \ref{fig:lc_rec_yes} shows the reconstructed light curves.

In order to compare the capability of our model to learn and encode physical parameters in the latent space, we compared two instances of the same model: 1) a cVAE network trained with light curves and labels, but without physical parameters as discussed in the previous section; 2) a similar network architecture (cVAE-P) that also includes physical parameters during training, concatenated to the label vector and provided to the encoder component of the network. After both models were trained, we establish a relation between the latent space and the physical parameters by fitting a multivariate regression between them. This allow us to select a given set of $T_{\rm eff}$, period, and absolute magnitude that are mapped to the latent space and later fed to the decoder in order to generate new light curves.

We evaluate three regression models\footnote{We used the \texttt{scikit-learn} \citep{scikit-learn} implementation for all three regression models}: linear, Random Forest (RF Regressor), and a basic Multi-Layer Perceptron (MLP). All three regressors are fitted using the same training set, the root-mean-square error (RMSE) of the validation set (20$\%$ of the total dataset) for each method is presented in Table \ref{tab:regr}. Both the linear and MLP regressions achieve similar RMSE, as expected, while the Random Forest regressor outperforms the others. Though the RF achieves a lower RMSE, tree-based regressors are restricted to predictions within the training set range. When comparing both generative models, with and without seeing physical parameters during training, RMSE values are not substantially different, but are consistently better for the model that includes physical parameters (cVAE-P).

\begin{deluxetable}{lcc}
\tablecaption{Latent-Physical space regression \label{tab:regr}}
\tablehead{\colhead{Generative Model} & \colhead{cVAE} & \colhead{cVAE-P}
}
\startdata
 Linear                      & 0.863 & 0.794 \\
 Random Forest               & {\bf 0.299} & {\bf 0.289} \\
 Multi-Layer Perceptron & 0.863 & 0.798 \\
\enddata
\tablecomments{Values correspond to the root-mean-square error for a validation set. Values in bold are the best achieved for each generative model.}
\end{deluxetable}

In order to keep the variational power of our generative model and avoid obtaining  the exact copy of the light curve when selecting a fixed vector of physical parameters, we added an extra ``dummy'' dimension to the physical space. Afterwards, the regression model is fitted with a collection of 100 repeated physical vectors per instance in the dataset that only differs in the value of the extra dimension, which is sampled from a uniform distribution. In the latent space, thanks to the variational architecture that encodes the parameters $\mu$ and $\sigma$ of the latent distributions, each latent vector is sampled 10 times from $\sim N(\mu, \sigma)$ for each instance of the dataset. This allows the model to generate slightly different time series for the same set of physical vectors, but keeping them consistent.

\subsection{Generating new Light Curves} \label{subsec:gen_lcs}

The process of generating a new light curve is described as follows: a vector of physical parameters is constructed given a set of values for effective temperature, absolute magnitude, and period; an extra dimension is added by sampling from $\sim U(0,1)$; this vector is projected into the latent space by means of the regression function, this returns a latent vector; this latent vector is tagged with the user-defined observed time-stamps calculated from the period and a zero-time value; this extended latent code is fed into the decoder network which returns a new phase-folded light curve, where phase values can be converted to time using the previously selected period and zero-time. With a model conditioned to the observed phases, it is possible to change the effective observational cadence of the light curve. This provides an opportunity to explicitly explore different observing cadences and how such cadences might impact the discovery and characterization potential of different variability classes.

\begin{figure*}[ht!]
\plotone{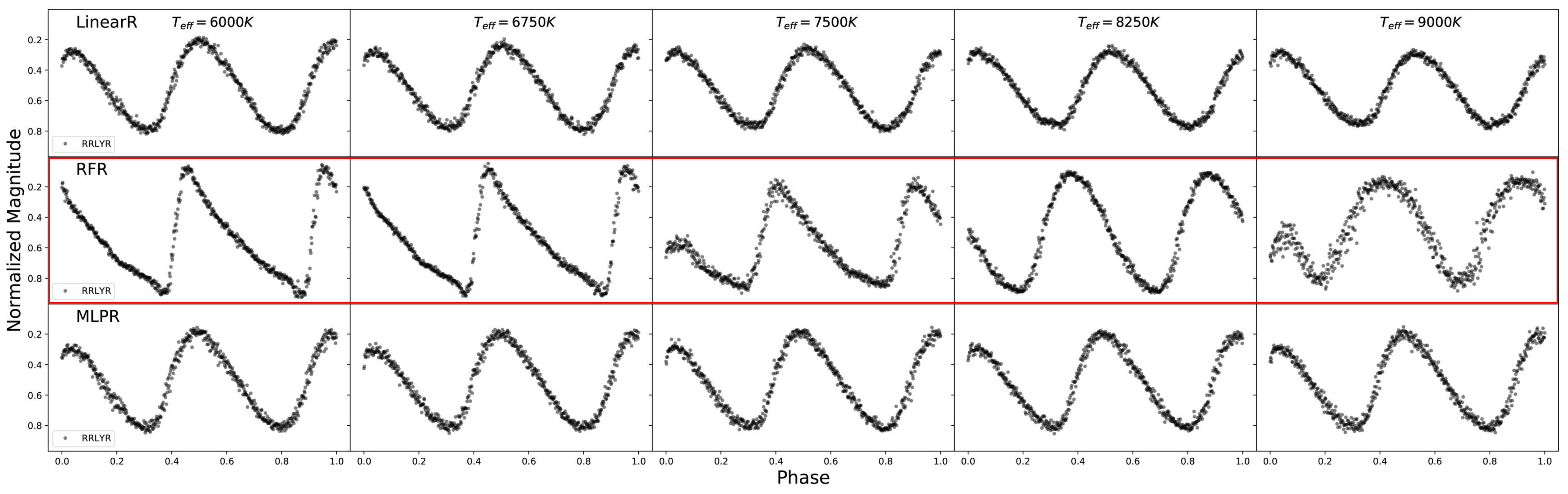}
\caption{Generated light curves of RR Lyrae stars as effective temperature $T_{\rm{eff}}$ increases (column direction). The three rows show the results from the three regression models: Linear, Random Forest, and Multi-Layer Perceptron. The RF regressor, second row, shows the better  representation along the temperature series for RR Lyrae variables, which agrees with its comparative RMSE performace.
\label{fig:lc_gen_temp_rrl}}
\end{figure*}

Figure \ref{fig:lc_gen_temp_rrl} presents a sequence of generated RR\,Lyrae light curves for different values of effective temperature as they increase in value. RR\,Lyrae light curves morph in shape with temperature increase, transitioning from a sawtooth shape characteristic of Bailey type \textit{ab} to a more sinusoidal shape typically found in hotter Bailey type \textit{c}. This change in light-curve shape is clearly shown when using the RF regressor (middle row), but minimal change in the shape for the other two regressors.

\subsection{Limitations and Future Explorations} \label{subsec:lim_future}

Incorporating a fixed number of physical parameters in learning the generative model limits its capacity to use others stellar parameters without retraining. On the other hand, the model trained only on light curves provides for the possibility of including {\it post facto} additional (physical) variables that were not explored in this study. This can done by fitting a new regression model to connect the latent space with the new space of physical parameters, avoiding the retraining of the VAE model. For instance, adding mettalicity, stellar mass, and surface gravity of the stars will provide a more complete generative model. The challenge here is obtaining a comprehensive training catalog of variable stars with  respective stellar parameters, across a variety of variability classes.

A future extension of this generative model would consist of finding a more complex and accurate connection between the latent space and the space of physical variables. Recent work has used Flow based models coupled to VAE models: \cite{2019arXiv191010046B} found, in an image-based domain, that by training a Normalizing Flow \citep[NF,][]{2015arXiv150505770J} to the encoded data distribution the sample quality improved when generating new images; both models provided competitive results with very little hyperparameter tuning. Moreover, the NF model also allows sampling from a true Normal distribution and then mapped to the latent distribution, which, while regularized to be Gaussian, in practice does not strictly follow a Gaussian distribution. Even more, a flow based model could capture better the covariance between physical parameters to find an even-more physically constrained mapping to the latent space. The later is specifically important for our purposes as each type of variable star tends to occupy a specific locus in the low-dimensional manifold of physical space. For instance, RR\,Lyrae are located in the intersection of the horizontal branch and the instability strip of the H-R diagram; this bounds the physically allowed values of effective temperature and luminosity to the \hbox{$[6000 - 7250]$\,K} range and $\sim 10^2 L_\odot$, respectively.

\section{Summary} \label{sec:summary}

To date, the most prominent uses of deep generative models have been in the image and spatial domains, with models that can tractably generate realistic landscapes, faces \citep{2019arXiv191204958K}, galaxies \citep{2019arXiv190912160D}, and dark matter distributions \citep{2019ComAC...6....1M}. Sequential data, primarily for natural language \citep{2017arXiv170510929R} and music \citep{engel2018gansynth}, has also been modelled with deep generative networks. However, previous to this work, we are not aware of the prior use of deep generative models in the astronomical time domain.

In this work,  we presented a deep generative model based on a variational autoencoder architecture that, after being trained with irregularly sampled and noisy light curve data, is able to reproduce and generate realistic periodic variable sources, such as RR Lyrae, eclipsing binaries, and Cepheids. This model includes an encoder module to extract relevant information from the light curves and auxiliary metadata (ie., physical parameters) and condense it into a low-level representation in latent space, and a decoder network that expands the latent code to reconstruction of the original time series. Both networks make use of temporal convolutional network layers followed by fully connected (dense) layers (Fig.~\ref{fig:vae_net}).

We trained this model with OGLE-III light curves and stellar parameters from the Gaia DR2 catalog. Our trained models are capable of recovering the distinctive characteristics of the light-curve shapes for eight different types of periodic variables. We present a preliminary version of the model trained only with light curves and a second model that includes physical parameters as ancillary inputs. For the first approach, the latent space only encodes the light-curve shape information, while for the second the latent space includes the information from physical parameters such as effective temperature, brightness (absolute magnitude), and period highlighting the correlation between the latent and physical space by means of multi-output regression. In that regard, we explored tree-based, linear, and multi-layer perceptron regression. Despite the limitations of tree-based aggregate learners to predict near the extrema of the target output variables, when using a root-mean-squared error loss, the Random Forest regressor showed the best result compared to simple linear and a one-hidden layer perceptron model.

With PELS-VAE, we introduce the methodology of generating new light curves by first selecting a vector of physical parameters which is projected into the latent space by means of a regression function. Afterwards, the latent vector is tagged with the desired observing timestamps and fed into the decoder network which creates a new light curve. The complete process of generating a batch of 100 new light curves on a modern CPU takes $\sim 1.3$ seconds, independent of the regression method, without parallelization.
The two generative models each present distinct advantages. The first model trained solely on the information from the phase-folded light curves is adjustable to include ancillary metadata (i.e., physical parameters) at a later stage without the need to re-train the model anew. The second model processes jointly the photometric observables and the metadata leading to a better mapping between the latent space and the physical space.
The exploration of highly-sophisticated models, such as autoregressive flows, that connect the space of physical parameters to the latent space, constitute a future work.

\acknowledgments

This research used the Exalearn computational cluster resource provided by the IT Division at the Lawrence Berkeley National Laboratory (Supported by the Director, Office of Science, Office of Basic Energy Sciences, of the U.S. Department of Energy under Contract No. DE--AC02--05CH11231).
JMP and JSB were partially supported by a Gordon and Betty Moore Foundation Data-Driven Discovery grant. ESA was supported by the National Science Foundation Graduate Research Fellowship under Grant No.\ DGE\,1752814.


\software{\texttt{numpy} \citep{2011CSE....13b..22V},
          \texttt{matplotlib} \citep{Hunter:2007},
          \texttt{jupyter} \citep{jupyter},
          \hbox{\texttt{pytorch}} \citep{paszke2017automatic},
          Weight \& Biases\footnote{\url{https://www.wandb.com/}},
          \texttt{scikit-learn} \citep{scikit-learn}, and
          \texttt{pandas} \citep{mckinney-proc-scipy-2010}
          }

\newpage
\appendix
\section{Gaia DR2 Cross-match Validation with OGLE-III} \label{apen:x-match}

The cross-match between the OGLE-III and Gaia DR2 catalogs use a $2 \arcsec$ search radius. This radius encloses all sources with proper motion (PM) up to 130 mas/year given the $\delta t = 15.5$\,yr difference between the effective epochs of the catalogs. The resulting cross-matched sources are constrained to $[-34,40]$ mas/year range (We found that a modest increase of this radius did not result in the addition of more matches). This supports the selection of a $2 \arcsec$ radius search without the need for compensation due to high PM sources.

To validate the resulting cross-matched sources, we first look for possible contaminating sources within the $2 \arcsec$ search radius. A total of $19545$ out of $34653$ sources do not have other neighbors within $2 \arcsec$ and each corresponding Gaia source has a calculated offset over $\delta t$ that is within the angular distance of the search. There are $15108$ sources that have multiple sources within $2 \arcsec$ of the OGLE source. Of these, $1113$ objects do not have neighbors within the angular distance of the OGLE source and the PM radius of the nearest Gaia source, and we accept these as valid matches. We acknowledge that this assumes that the nearest-neighbor match is correct in order to rely on Gaia PM values, which can be supported by the analysis presented above. Only $6$ of $15108$ objects have a cross-match angular distance larger that their own PM radius value. We visually inspect those cases using the Aladin Sky Atlas platform \citep{2000A&AS..143...33B}. Five of these are valid cross-matches, while one source failed a by-eye confirmation of the Gaia DR2 Catalog overlaid on Pan-STARRS images as the baseline.

The remaining $13989$ sources had more than one object within the PM radius of the nearest star. Of these, $3,835$ are listed with variable star classifications in the Gaia Variability Catalog \citep{2019A&A...625A..97R}. In particular, $3732$ have matching variability classification (without considering sub-types) between the OGLE and Gaia Variability catalogs. The remainder ($103$) have mismatched classification labels, from which 89 are sources either classified as RR Lyrae or Cepheids by one or the other catalog with measured periods of $< 1$ day. A similar result was obtained for $9$ of $103$ objects, which are labeled as eclipsing binary systems by OGLE but have a different label in Gaia catalogs. This confusion is expected due to the sparse amount of observations ($\sim$30 data points) and uneven windowing present in Gaia light curves when compared to the denser and longer-baseline OGLE time-series. Therefore, after a visual inspection of light curves from both OGLE and Gaia we confirmed 98 cross-matches and adopted the OGLE classification. We discarded the remaining 5 sources due to a catastrophic mismatch in their classification type and visual inspection of the lightcurves between the two catalogs. A further $461$ of $13989$ sources are flagged as variables in the Gaia DR2 catalog but have no assigned variability subtype. For this subsample, we check the Gaia colors ($G_{\rm BP} - G_{\rm RP}$) and effective temperatures ($T_{e\rm ff}$) against the corresponding ranges per-class for the $19545$ confirmed objects, informed by the known value ranges available in the literature \citep{2015pust.book.....C}, allowing us to validate the 461 sources as likely correct matches.

\begin{figure*}[ht!]
\plotone{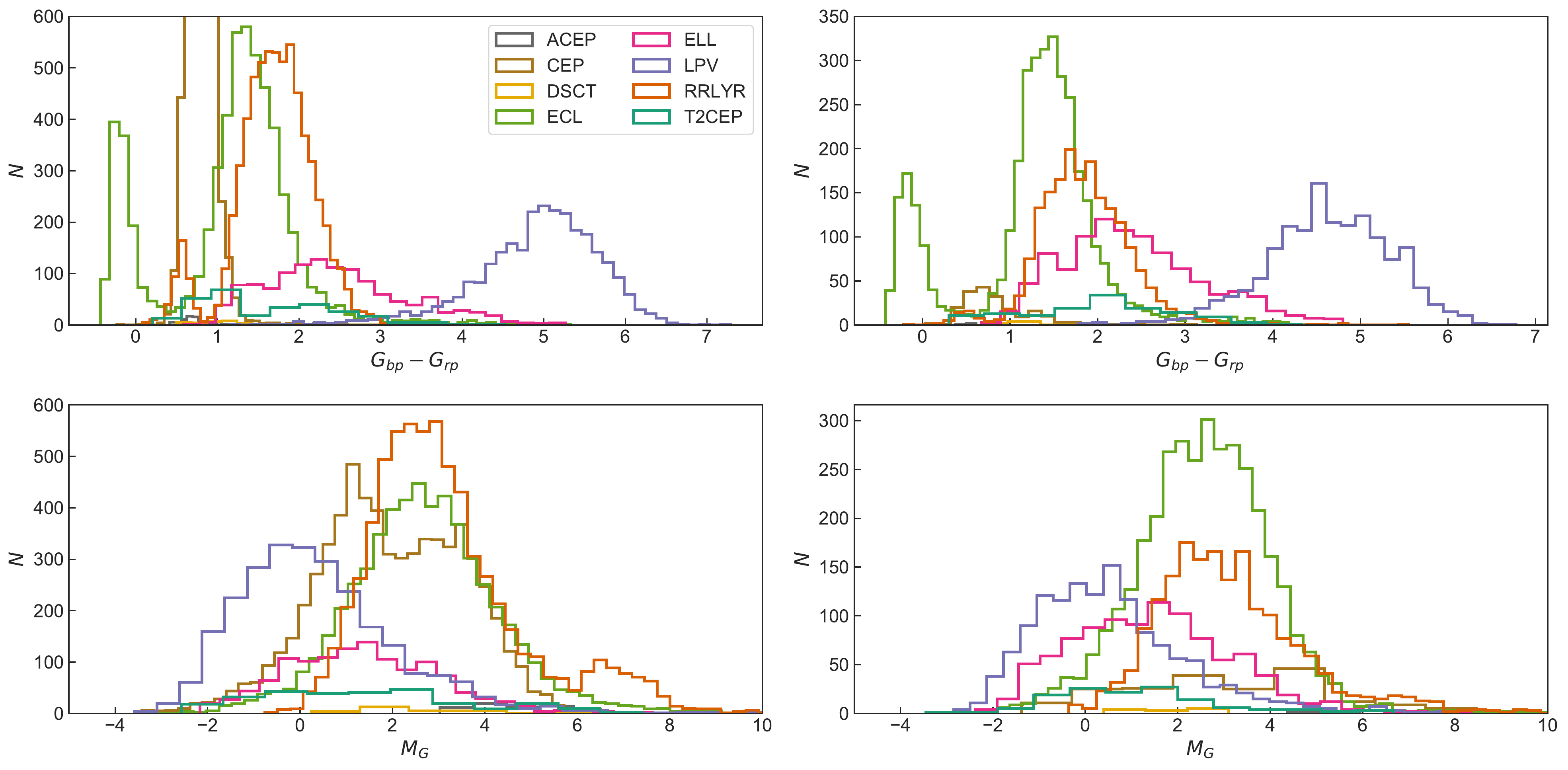}
\caption{Distributions of Gaia color ($G_{\rm BP} - G_{\rm RP}$, upper row) and absolute \textit{g}-band magnitudes (M$_{\rm G}$, lower row) color codded by variability classes. The left two panels show the confirmed cross-matches (sources with only one match within PM radius and with matching variability class), while the right panels show the sources validated by CMD comparisons (last group in Table \ref{tab:xmatch_counts}).
\label{fig:lum_color_dist}}
\end{figure*}

Finally, the remaining $9693$ sources, from the previous count of $13989$ variable stars, have more than one neighbor within their PM radius and their nearest-neighbor match has no variability information provided by Gaia. To validate this subset, we first filtered following the same temperature and color criteria described in the previous paragraph. However, without additional information on the variability of these sources from Gaia, we further analyzed these sources by inspecting the position of the nearest Gaia source in the color magnitude diagram (CMD), using the confirmed sample and the locus for known pulsating variables \citep{2019A&A...623A.110G} as a ground truth comparison. There are $8374$ of $9693$ sources that have $G_{\rm BP}$ and $G_{\rm RP}$ measurements in Gaia DR2, which we combined with the estimated distances from (\citealt{ 2018AJ....156...58B}; hereafter the Bailer-Jones catalog) in order to account for known issues with Gaia parallax measurements in crowded areas like the plane of the galaxy. To avoid color and magnitude degeneracies with possible mismatches, we remove 775 sources within the main sequence region that could contaminate our sample, validating 7620 nearest neighbor cross-matches. We reject the 465 objects with three or more sources in the search radius, except for the two objects whose second-closest neighbor has matching Gaia and OGLE classes. The $853$ objects that remain are missing $G_{\rm BP}$ and $G_{\rm RP}$ for the nearest-neighbor match, but the only other star in the cross-match radius had measured colors. There are $151$ of $853$ objects that do not have estimated distances in the Bailer-Jones catalog and we reject these cross-matches. Since none of the objects in our catalog with confirmed Gaia classes were missing Gaia colors, we placed the second-closest Gaia match on the CMD for the remaining $701$ sources, and found that 539 sources were in the correct place on the CMD for their OGLE classes. We reject the 162 sources that were degenerate with the main-sequence or had Gaia colors that were too blue for the OGLE class. Lastly, after all validation steps, we consolidate our dataset with $33114$ valid cross-matches. There are $32573$ of these matches that were with the nearest-neighbor and $541$ matches were with the second nearest-neighbor, all within the PM radius of the Gaia source. We make the OGLE-III/Gaia DR2 cross-match publicly available on Zenodo {\bf link TBD}.

\begin{deluxetable*}{ll}
\tablecaption{Cross-match Validation Summary
\label{tab:xmatch_counts}}
\tablewidth{0pt}
\tablehead{
\colhead{Count} & \colhead{Description}
}
\startdata
 34,653 & Size of 2$\arcsec$ cross-match between OGLE and Gaia\\
 19,545 & Objects that only have 1 match in 2$\arcsec$ radius\\
 1,113 & Objects that have $> 1$ match in 2$\arcsec$ radius but only 1 match in PM radius of nearest neighbor\\
 5 & Objects with a cross-match angular distance outside of the nearest neighbor's PM radius that pass visual inspection\\
 \textcolor{dark_red}{1} & \textcolor{dark_red}{Object with an angular distance outside of the PM radius that doesn't pass visual inspection} \vspace{10 px} \\
 \hline
 \multicolumn{2}{c}{4,296 Objects with $> 1$ match in both 2$\arcsec$ radius and nearest neighbor PM radius and variability information} \vspace{2px} \\
 \hline
 3,732 & Objects with the same variability class in OGLE and Gaia\\
 461 & Objects flagged as variable with no assigned class, similar $T_{\rm eff}$ and colors to 19,545 confirmed \\
 98 & Objects with adjacent variability classes and visually similar lightcurves \\
 \textcolor{dark_red}{5} & \textcolor{dark_red}{Objects with fatally different classes and visually dissimilar lightcurves} \vspace{10 px} \\
 \hline
 \multicolumn{2}{c}{9,693 Objects with $> 1$ match in both 2$\arcsec$ radius and nearest neighbor PM radius and no variability information} \vspace{2px} \\
 \hline
  7,620 & Objects that have $G_{\rm BP}$ and $G_{\rm RP}$ measurements in Gaia DR2, validated by CMD placement \\
  \textcolor{purple}{539} & \textcolor{purple}{Objects whose nearest neighbor lacks $G_{\rm BP}$ and $G_{\rm RP}$ measurements, second-closest neighbor validated by CMD placement} \\
  \textcolor{purple}{2} & \textcolor{purple}{Objects whose second-closest neighbor has the same variability class in OGLE and Gaia} \\
  \textcolor{dark_red}{755} & \textcolor{dark_red}{Objects that have $G_{\rm BP}$ and $G_{\rm RP}$ measurements in Gaia DR2, degenerate with the main sequence}\\
  \textcolor{dark_red}{465} & \textcolor{dark_red}{Objects with three or more Gaia sources within the nearest-neighbor PM radius}\\
  \textcolor{dark_red}{162} & \textcolor{dark_red}{Objects whose second-closest neighbor is degenerate with the main sequence or Gaia colors too blue for the OGLE class}\\
  \textcolor{dark_red}{151} & \textcolor{dark_red}{Objects do not have inferred distances in the Bailer-Jones catalog for any stars in the cross-match radius}\\
   & \\
\enddata
\tablecomments{Lines in red were dropped from the cross-match as definite or possible mismatches. For Lines in purple, the second-closest neighbor provided the correct match.}
\end{deluxetable*}

\bibliography{main.bbl}

\begin{thebibliography}{}
\expandafter\ifx\csname natexlab\endcsname\relax\def\natexlab#1{#1}\fi
\providecommand{\url}[1]{\href{#1}{#1}}
\providecommand{\dodoi}[1]{doi:~\href{http://doi.org/#1}{\nolinkurl{#1}}}
\providecommand{\doeprint}[1]{\href{http://ascl.net/#1}{\nolinkurl{http://ascl.net/#1}}}
\providecommand{\doarXiv}[1]{\href{https://arxiv.org/abs/#1}{\nolinkurl{https://arxiv.org/abs/#1}}}

\bibitem[{{Aguirre} {et~al.}(2019){Aguirre}, {Pichara}, \&
  {Becker}}]{2019MNRAS.482.5078A}
{Aguirre}, C., {Pichara}, K., \& {Becker}, I. 2019, \mnras, 482, 5078,
  \dodoi{10.1093/mnras/sty2836}

\bibitem[{{Andrae} {et~al.}(2018){Andrae}, {Fouesneau}, {Creevey}, {Ordenovic},
  {Mary}, {Burlacu}, {Chaoul}, {Jean-Antoine-Piccolo}, {Kordopatis}, {Korn},
  {Lebreton}, {Panem}, {Pichon}, {Th{\'e}venin}, {Walmsley}, \&
  {Bailer-Jones}}]{2018A&A...616A...8A}
{Andrae}, R., {Fouesneau}, M., {Creevey}, O., {et~al.} 2018, \aap, 616, A8,
  \dodoi{10.1051/0004-6361/201732516}

\bibitem[{{Bachelet} {et~al.}(2017){Bachelet}, {Norbury}, {Bozza}, \&
  {Street}}]{2017AJ....154..203B}
{Bachelet}, E., {Norbury}, M., {Bozza}, V., \& {Street}, R. 2017, \aj, 154,
  203, \dodoi{10.3847/1538-3881/aa911c}

\bibitem[{{Bai} {et~al.}(2018){Bai}, {Zico Kolter}, \&
  {Koltun}}]{2018arXiv180301271B}
{Bai}, S., {Zico Kolter}, J., \& {Koltun}, V. 2018, arXiv e-prints,
  arXiv:1803.01271.
\newblock \doarXiv{1803.01271}

\bibitem[{{Bailer-Jones} {et~al.}(2018){Bailer-Jones}, {Rybizki}, {Fouesneau},
  {Mantelet}, \& {Andrae}}]{2018AJ....156...58B}
{Bailer-Jones}, C.~A.~L., {Rybizki}, J., {Fouesneau}, M., {Mantelet}, G., \&
  {Andrae}, R. 2018, \aj, 156, 58, \dodoi{10.3847/1538-3881/aacb21}

\bibitem[{Bellm {et~al.}(2018)Bellm, Kulkarni, Graham, Dekany, Smith, Riddle,
  Masci, Helou, Prince, Adams, Barbarino, Barlow, Bauer, Beck, Belicki, Biswas,
  Blagorodnova, Bodewits, Bolin, Brinnel, Brooke, Bue, Bulla, Burruss, Cenko,
  Chang, Connolly, Coughlin, Cromer, Cunningham, De, Delacroix, Desai, Duev,
  Eadie, Farnham, Feeney, Feindt, Flynn, Franckowiak, Frederick, Fremling,
  Gal-Yam, Gezari, Giomi, Goldstein, Golkhou, Goobar, Groom, Hacopians, Hale,
  Henning, Ho, Hover, Howell, Hung, Huppenkothen, Imel, Ip, Ivezi{\'{c}},
  Jackson, Jones, Juric, Kasliwal, Kaspi, Kaye, Kelley, Kowalski, Kramer,
  Kupfer, Landry, Laher, Lee, Lin, Lin, Lunnan, Giomi, Mahabal, Mao, Miller,
  Monkewitz, Murphy, Ngeow, Nordin, Nugent, Ofek, Patterson, Penprase, Porter,
  Rauch, Rebbapragada, Reiley, Rigault, Rodriguez, van Roestel, Rusholme, van
  Santen, Schulze, Shupe, Singer, Soumagnac, Stein, Surace, Sollerman, Szkody,
  Taddia, Terek, Sistine, van Velzen, Vestrand, Walters, Ward, Ye, Yu, Yan, \&
  Zolkower}]{Bellm_2018}
Bellm, E.~C., Kulkarni, S.~R., Graham, M.~J., {et~al.} 2018, Publications of
  the Astronomical Society of the Pacific, 131, 018002,
  \dodoi{10.1088/1538-3873/aaecbe}

\bibitem[{{Benavente} {et~al.}(2017){Benavente}, {Protopapas}, \&
  {Pichara}}]{2017ApJ...845..147B}
{Benavente}, P., {Protopapas}, P., \& {Pichara}, K. 2017, \apj, 845, 147,
  \dodoi{10.3847/1538-4357/aa7f2d}

\bibitem[{{Bengio} {et~al.}(1994){Bengio}, {Simard}, \& {Frasconi}}]{279181}
{Bengio}, Y., {Simard}, P., \& {Frasconi}, P. 1994, IEEE Transactions on Neural
  Networks, 5, 157

\bibitem[{{Bloom} {et~al.}(2012){Bloom}, {Richards}, {Nugent}, {Quimby},
  {Kasliwal}, {Starr}, {Poznanski}, {Ofek}, {Cenko}, {Butler}, {Kulkarni},
  {Gal-Yam}, \& {Law}}]{Bloom12}
{Bloom}, J.~S., {Richards}, J.~W., {Nugent}, P.~E., {et~al.} 2012, Publications
  of the Astronomical Society of the Pacific, 124, 1175, \dodoi{10.1086/668468}

\bibitem[{{B{\"o}hm} {et~al.}(2019){B{\"o}hm}, {Lanusse}, \&
  {Seljak}}]{2019arXiv191010046B}
{B{\"o}hm}, V., {Lanusse}, F., \& {Seljak}, U. 2019, arXiv e-prints,
  arXiv:1910.10046.
\newblock \doarXiv{1910.10046}

\bibitem[{{Bonnarel} {et~al.}(2000){Bonnarel}, {Fernique}, {Bienaym{\'e}},
  {Egret}, {Genova}, {Louys}, {Ochsenbein}, {Wenger}, \&
  {Bartlett}}]{2000A&AS..143...33B}
{Bonnarel}, F., {Fernique}, P., {Bienaym{\'e}}, O., {et~al.} 2000, \aaps, 143,
  33, \dodoi{10.1051/aas:2000331}

\bibitem[{{Boone}(2019)}]{2019AJ....158..257B}
{Boone}, K. 2019, \aj, 158, 257, \dodoi{10.3847/1538-3881/ab5182}

\bibitem[{{Burgess} {et~al.}(2018){Burgess}, {Higgins}, {Pal}, {Matthey},
  {Watters}, {Desjardins}, \& {Lerchner}}]{2018arXiv180403599B}
{Burgess}, C.~P., {Higgins}, I., {Pal}, A., {et~al.} 2018, arXiv e-prints,
  arXiv:1804.03599.
\newblock \doarXiv{1804.03599}

\bibitem[{{Cabrera-Vives} {et~al.}(2017){Cabrera-Vives}, {Reyes},
  {F{\"o}rster}, {Est{\'e}vez}, \& {Maureira}}]{Cabrera17}
{Cabrera-Vives}, G., {Reyes}, I., {F{\"o}rster}, F., {Est{\'e}vez}, P.~A., \&
  {Maureira}, J.-C. 2017, \apj, 836, 97, \dodoi{10.3847/1538-4357/836/1/97}

\bibitem[{{Carrasco-Davis} {et~al.}(2019){Carrasco-Davis}, {Cabrera-Vives},
  {F{\"o}rster}, {Est{\'e}vez}, {Huijse}, {Protopapas}, {Reyes},
  {Mart{\'\i}nez-Palomera}, \& {Donoso}}]{2019PASP..131j8006C}
{Carrasco-Davis}, R., {Cabrera-Vives}, G., {F{\"o}rster}, F., {et~al.} 2019,
  \pasp, 131, 108006, \dodoi{10.1088/1538-3873/aaef12}

\bibitem[{{Catelan} \& {Smith}(2015)}]{2015pust.book.....C}
{Catelan}, M., \& {Smith}, H.~A. 2015, {Pulsating Stars}

\bibitem[{{Cho} {et~al.}(2014){Cho}, {van Merrienboer}, {Gulcehre}, {Bahdanau},
  {Bougares}, {Schwenk}, \& {Bengio}}]{2014arXiv1406.1078C}
{Cho}, K., {van Merrienboer}, B., {Gulcehre}, C., {et~al.} 2014, arXiv
  e-prints, arXiv:1406.1078.
\newblock \doarXiv{1406.1078}

\bibitem[{{Dia} {et~al.}(2019){Dia}, {Savary}, {Melchior}, \&
  {Courbin}}]{2019arXiv190912160D}
{Dia}, M., {Savary}, E., {Melchior}, M., \& {Courbin}, F. 2019, arXiv e-prints,
  arXiv:1909.12160.
\newblock \doarXiv{1909.12160}

\bibitem[{{Dieleman} {et~al.}(2015){Dieleman}, {Willett}, \&
  {Dambre}}]{Dieleman15}
{Dieleman}, S., {Willett}, K.~W., \& {Dambre}, J. 2015, \mnras, 450, 1441,
  \dodoi{10.1093/mnras/stv632}

\bibitem[{{Drake} {et~al.}(2013){Drake}, {Catelan}, {Djorgovski}, {Torrealba},
  {Graham}, {Belokurov}, {Koposov}, {Mahabal}, {Prieto}, {Donalek}, {Williams},
  {Larson}, {Christensen}, \& {Beshore}}]{2013ApJ...763...32D}
{Drake}, A.~J., {Catelan}, M., {Djorgovski}, S.~G., {et~al.} 2013, \apj, 763,
  32, \dodoi{10.1088/0004-637X/763/1/32}

\bibitem[{Engel {et~al.}(2019)Engel, Agrawal, Chen, Gulrajani, Donahue, \&
  Roberts}]{engel2018gansynth}
Engel, J., Agrawal, K.~K., Chen, S., {et~al.} 2019, in International Conference
  on Learning Representations.
\newblock \url{https://openreview.net/forum?id=H1xQVn09FX}

\bibitem[{{F{\"o}rster} {et~al.}(2016){F{\"o}rster}, {Maureira}, {San
  Mart{\'\i}n}, {Hamuy}, {Mart{\'\i}nez}, {Huijse}, {Cabrera}, {Galbany}, {de
  Jaeger}, {Gonz{\'a}lez─Gait{\'a}n}, {Anderson}, {Kunkarayakti}, {Pignata},
  {Bufano}, {Litt{\'\i}n}, {Olivares}, {Medina}, {Smith}, {Vivas},
  {Est{\'e}vez}, {Mu{\~n}oz}, \& {Vera}}]{2016ApJ...832..155F}
{F{\"o}rster}, F., {Maureira}, J.~C., {San Mart{\'\i}n}, J., {et~al.} 2016,
  \apj, 832, 155, \dodoi{10.3847/0004-637X/832/2/155}

\bibitem[{{Gabbard} {et~al.}(2019){Gabbard}, {Messenger}, {Heng}, {Tonolini},
  \& {Murray-Smith}}]{2019arXiv190906296G}
{Gabbard}, H., {Messenger}, C., {Heng}, I.~S., {Tonolini}, F., \&
  {Murray-Smith}, R. 2019, arXiv e-prints, arXiv:1909.06296.
\newblock \doarXiv{1909.06296}

\bibitem[{{Gaia Collaboration} {et~al.}(2016){Gaia Collaboration}, {Prusti},
  {de Bruijne}, {Brown}, {Vallenari}, {Babusiaux}, {Bailer-Jones}, {Bastian},
  {Biermann}, {Evans}, {Eyer}, {Jansen}, {Jordi}, {Klioner}, {Lammers},
  {Lindegren}, {Luri}, {Mignard}, {Milligan}, {Panem}, {Poinsignon},
  {Pourbaix}, {Randich}, {Sarri}, {Sartoretti}, {Siddiqui}, {Soubiran},
  {Valette}, {van Leeuwen}, {Walton}, {Aerts}, {Arenou}, {Cropper}, {Drimmel},
  {H{\o}g}, {Katz}, {Lattanzi}, {O'Mullane}, {Grebel}, {Holland}, {Huc},
  {Passot}, {Bramante}, {Cacciari}, {Casta{\~n}eda}, {Chaoul}, {Cheek}, {De
  Angeli}, {Fabricius}, {Guerra}, {Hern{\'a}ndez}, {Jean-Antoine-Piccolo},
  {Masana}, {Messineo}, {Mowlavi}, {Nienartowicz}, {Ord{\'o}{\~n}ez-Blanco},
  {Panuzzo}, {Portell}, {Richards}, {Riello}, {Seabroke}, {Tanga},
  {Th{\'e}venin}, {Torra}, {Els}, {Gracia-Abril}, {Comoretto},
  {Garcia-Reinaldos}, {Lock}, {Mercier}, {Altmann}, {Andrae}, {Astraatmadja},
  {Bellas-Velidis}, {Benson}, {Berthier}, {Blomme}, {Busso}, {Carry},
  {Cellino}, {Clementini}, {Cowell}, {Creevey}, {Cuypers}, {Davidson}, {De
  Ridder}, {de Torres}, {Delchambre}, {Dell'Oro}, {Ducourant}, {Fr{\'e}mat},
  {Garc{\'\i}a-Torres}, {Gosset}, {Halbwachs}, {Hambly}, {Harrison}, {Hauser},
  {Hestroffer}, {Hodgkin}, {Huckle}, {Hutton}, {Jasniewicz}, {Jordan},
  {Kontizas}, {Korn}, {Lanzafame}, {Manteiga}, {Moitinho}, {Muinonen},
  {Osinde}, {Pancino}, {Pauwels}, {Petit}, {Recio-Blanco}, {Robin}, {Sarro},
  {Siopis}, {Smith}, {Smith}, {Sozzetti}, {Thuillot}, {van Reeven}, {Viala},
  {Abbas}, {Abreu Aramburu}, {Accart}, {Aguado}, {Allan}, {Allasia},
  {Altavilla}, {{\'A}lvarez}, {Alves}, {Anderson}, {Andrei}, {Anglada Varela},
  {Antiche}, {Antoja}, {Ant{\'o}n}, {Arcay}, {Atzei}, {Ayache}, {Bach},
  {Baker}, {Balaguer-N{\'u}{\~n}ez}, {Barache}, {Barata}, {Barbier}, {Barblan},
  {Baroni}, {Barrado y Navascu{\'e}s}, {Barros}, {Barstow}, {Becciani},
  {Bellazzini}, {Bellei}, {Bello Garc{\'\i}a}, {Belokurov}, {Bendjoya},
  {Berihuete}, {Bianchi}, {Bienaym{\'e}}, {Billebaud}, {Blagorodnova},
  {Blanco-Cuaresma}, {Boch}, {Bombrun}, {Borrachero}, {Bouquillon}, {Bourda},
  {Bouy}, {Bragaglia}, {Breddels}, {Brouillet}, {Br{\"u}semeister},
  {Bucciarelli}, {Budnik}, {Burgess}, {Burgon}, {Burlacu}, {Busonero}, {Buzzi},
  {Caffau}, {Cambras}, {Campbell}, {Cancelliere}, {Cantat-Gaudin}, {Carlucci},
  {Carrasco}, {Castellani}, {Charlot}, {Charnas}, {Charvet}, {Chassat},
  {Chiavassa}, {Clotet}, {Cocozza}, {Collins}, {Collins}, {Costigan}, {Crifo},
  {Cross}, {Crosta}, {Crowley}, {Dafonte}, {Damerdji}, {Dapergolas}, {David},
  {David}, {De Cat}, {de Felice}, {de Laverny}, {De Luise}, {De March}, {de
  Martino}, {de Souza}, {Debosscher}, {del Pozo}, {Delbo}, {Delgado},
  {Delgado}, {di Marco}, {Di Matteo}, {Diakite}, {Distefano}, {Dolding}, {Dos
  Anjos}, {Drazinos}, {Dur{\'a}n}, {Dzigan}, {Ecale}, {Edvardsson}, {Enke},
  {Erdmann}, {Escolar}, {Espina}, {Evans}, {Eynard Bontemps}, {Fabre},
  {Fabrizio}, {Faigler}, {Falc{\~a}o}, {Farr{\`a}s Casas}, {Faye}, {Federici},
  {Fedorets}, {Fern{\'a}ndez-Hern{\'a}ndez}, {Fernique}, {Fienga}, {Figueras},
  {Filippi}, {Findeisen}, {Fonti}, {Fouesneau}, {Fraile}, {Fraser}, {Fuchs},
  {Furnell}, {Gai}, {Galleti}, {Galluccio}, {Garabato}, {Garc{\'\i}a-Sedano},
  {Gar{\'e}}, {Garofalo}, {Garralda}, {Gavras}, {Gerssen}, {Geyer}, {Gilmore},
  {Girona}, {Giuffrida}, {Gomes}, {Gonz{\'a}lez-Marcos},
  {Gonz{\'a}lez-N{\'u}{\~n}ez}, {Gonz{\'a}lez-Vidal}, {Granvik}, {Guerrier},
  {Guillout}, {Guiraud}, {G{\'u}rpide}, {Guti{\'e}rrez-S{\'a}nchez}, {Guy},
  {Haigron}, {Hatzidimitriou}, {Haywood}, {Heiter}, {Helmi}, {Hobbs},
  {Hofmann}, {Holl}, {Holland }, {Hunt}, {Hypki}, {Icardi}, {Irwin}, {Jevardat
  de Fombelle}, {Jofr{\'e}}, {Jonker}, {Jorissen}, {Julbe}, {Karampelas},
  {Kochoska}, {Kohley}, {Kolenberg}, {Kontizas}, {Koposov}, {Kordopatis},
  {Koubsky}, {Kowalczyk}, {Krone-Martins}, {Kudryashova}, {Kull}, {Bachchan},
  {Lacoste-Seris}, {Lanza}, {Lavigne}, {Le Poncin-Lafitte}, {Lebreton},
  {Lebzelter}, {Leccia}, {Leclerc}, {Lecoeur-Taibi}, {Lemaitre}, {Lenhardt},
  {Leroux}, {Liao}, {Licata}, {Lindstr{\o}m}, {Lister}, {Livanou}, {Lobel},
  {L{\"o}ffler}, {L{\'o}pez}, {Lopez-Lozano}, {Lorenz}, {Loureiro},
  {MacDonald}, {Magalh{\~a}es Fernandes}, {Managau}, {Mann}, {Mantelet},
  {Marchal}, {Marchant}, {Marconi}, {Marie}, {Marinoni}, {Marrese},
  {Marschalk{\'o}}, {Marshall}, {Mart{\'\i}n-Fleitas}, {Martino}, {Mary},
  {Matijevi{\v{c}}}, {Mazeh}, {McMillan}, {Messina}, {Mestre}, {Michalik},
  {Millar}, {Miranda}, {Molina}, {Molinaro}, {Molinaro}, {Moln{\'a}r},
  {Moniez}, {Montegriffo}, {Monteiro}, {Mor}, {Mora}, {Morbidelli}, {Morel},
  {Morgenthaler}, {Morley}, {Morris}, {Mulone}, {Muraveva}, {Musella},
  {Narbonne}, {Nelemans}, {Nicastro}, {Noval}, {Ord{\'e}novic},
  {Ordieres-Mer{\'e}}, {Osborne}, {Pagani}, {Pagano}, {Pailler}, {Palacin},
  {Palaversa}, {Parsons}, {Paulsen}, {Pecoraro}, {Pedrosa}, {Pentik{\"a}inen},
  {Pereira}, {Pichon}, {Piersimoni}, {Pineau}, {Plachy}, {Plum}, {Poujoulet},
  {Pr{\v{s}}a}, {Pulone}, {Ragaini}, {Rago}, {Rambaux}, {Ramos-Lerate},
  {Ranalli}, {Rauw}, {Read}, {Regibo}, {Renk}, {Reyl{\'e}}, {Ribeiro},
  {Rimoldini}, {Ripepi}, {Riva}, {Rixon}, {Roelens}, {Romero-G{\'o}mez},
  {Rowell}, {Royer}, {Rudolph}, {Ruiz-Dern}, {Sadowski}, {Sagrist{\`a}
  Sell{\'e}s}, {Sahlmann}, {Salgado}, {Salguero}, {Sarasso}, {Savietto},
  {Schnorhk}, {Schultheis}, {Sciacca}, {Segol}, {Segovia}, {Segransan},
  {Serpell}, {Shih}, {Smareglia}, {Smart}, {Smith}, {Solano}, {Solitro},
  {Sordo}, {Soria Nieto}, {Souchay}, {Spagna}, {Spoto}, {Stampa}, {Steele},
  {Steidelm{\"u}ller}, {Stephenson}, {Stoev}, {Suess}, {S{\"u}veges}, {Surdej},
  {Szabados}, {Szegedi-Elek}, {Tapiador}, {Taris}, {Tauran}, {Taylor},
  {Teixeira}, {Terrett}, {Tingley}, {Trager}, {Turon}, {Ulla}, {Utrilla},
  {Valentini}, {van Elteren}, {Van Hemelryck}, {van Leeuwen}, {Varadi},
  {Vecchiato}, {Veljanoski}, {Via}, {Vicente}, {Vogt}, {Voss}, {Votruba},
  {Voutsinas}, {Walmsley}, {Weiler}, {Weingrill}, {Werner}, {Wevers},
  {Whitehead}, {Wyrzykowski}, {Yoldas}, {{\v{Z}}erjal}, {Zucker}, {Zurbach},
  {Zwitter}, {Alecu}, {Allen}, {Allende Prieto}, {Amorim},
  {Anglada-Escud{\'e}}, {Arsenijevic}, {Azaz}, {Balm}, {Beck}, {Bernstein},
  {Bigot}, {Bijaoui}, {Blasco}, {Bonfigli}, {Bono}, {Boudreault}, {Bressan},
  {Brown}, {Brunet}, {Bunclark}, {Buonanno}, {Butkevich}, {Carret}, {Carrion},
  {Chemin}, {Ch{\'e}reau}, {Corcione}, {Darmigny}, {de Boer}, {de Teodoro}, {de
  Zeeuw}, {Delle Luche}, {Domingues}, {Dubath}, {Fodor}, {Fr{\'e}zouls},
  {Fries}, {Fustes}, {Fyfe}, {Gallardo}, {Gallegos}, {Gardiol}, {Gebran},
  {Gomboc}, {G{\'o}mez}, {Grux}, {Gueguen}, {Heyrovsky}, {Hoar}, {Iannicola},
  {Isasi Parache}, {Janotto}, {Joliet}, {Jonckheere}, {Keil}, {Kim},
  {Klagyivik}, {Klar}, {Knude}, {Kochukhov}, {Kolka}, {Kos}, {Kutka}, {Lainey},
  {LeBouquin}, {Liu}, {Loreggia}, {Makarov}, {Marseille}, {Martayan},
  {Martinez-Rubi}, {Massart}, {Meynadier}, {Mignot}, {Munari}, {Nguyen},
  {Nordlander}, {Ocvirk}, {O'Flaherty}, {Olias Sanz}, {Ortiz}, {Osorio},
  {Oszkiewicz}, {Ouzounis}, {Palmer}, {Park}, {Pasquato}, {Peltzer}, {Peralta},
  {P{\'e}turaud}, {Pieniluoma}, {Pigozzi}, {Poels}, {Prat}, {Prod'homme},
  {Raison}, {Rebordao}, {Risquez}, {Rocca-Volmerange}, {Rosen}, {Ruiz-Fuertes},
  {Russo}, {Sembay}, {Serraller Vizcaino}, {Short}, {Siebert}, {Silva},
  {Sinachopoulos}, {Slezak}, {Soffel}, {Sosnowska}, {Strai{\v{z}}ys}, {ter
  Linden}, {Terrell}, {Theil}, {Tiede}, {Troisi}, {Tsalmantza}, {Tur},
  {Vaccari}, {Vachier}, {Valles}, {Van Hamme}, {Veltz}, {Virtanen}, {Wallut},
  {Wichmann}, {Wilkinson}, {Ziaeepour}, \& {Zschocke}}]{2016A&A...595A...1G}
{Gaia Collaboration}, {Prusti}, T., {de Bruijne}, J.~H.~J., {et~al.} 2016,
  \aap, 595, A1, \dodoi{10.1051/0004-6361/201629272}

\bibitem[{{Gaia Collaboration} {et~al.}(2018){Gaia Collaboration}, {Brown},
  {Vallenari}, {Prusti}, {de Bruijne}, {Babusiaux}, {Bailer-Jones}, {Biermann},
  {Evans}, {Eyer}, {Jansen}, {Jordi}, {Klioner}, {Lammers}, {Lindegren},
  {Luri}, {Mignard}, {Panem}, {Pourbaix}, {Randich}, {Sartoretti}, {Siddiqui},
  {Soubiran}, {van Leeuwen}, {Walton}, {Arenou}, {Bastian}, {Cropper},
  {Drimmel}, {Katz}, {Lattanzi}, {Bakker}, {Cacciari}, {Casta{\~n}eda},
  {Chaoul}, {Cheek}, {De Angeli}, {Fabricius}, {Guerra}, {Holl}, {Masana},
  {Messineo}, {Mowlavi}, {Nienartowicz}, {Panuzzo}, {Portell}, {Riello},
  {Seabroke}, {Tanga}, {Th{\'e}venin}, {Gracia-Abril}, {Comoretto},
  {Garcia-Reinaldos}, {Teyssier}, {Altmann}, {Andrae}, {Audard},
  {Bellas-Velidis}, {Benson}, {Berthier}, {Blomme}, {Burgess}, {Busso},
  {Carry}, {Cellino}, {Clementini}, {Clotet}, {Creevey}, {Davidson}, {De
  Ridder}, {Delchambre}, {Dell'Oro}, {Ducourant},
  {Fern{\'a}ndez-Hern{\'a}ndez}, {Fouesneau}, {Fr{\'e}mat}, {Galluccio},
  {Garc{\'\i}a-Torres}, {Gonz{\'a}lez-N{\'u}{\~n}ez}, {Gonz{\'a}lez-Vidal},
  {Gosset}, {Guy}, {Halbwachs}, {Hambly}, {Harrison}, {Hern{\'a}ndez},
  {Hestroffer}, {Hodgkin}, {Hutton}, {Jasniewicz}, {Jean-Antoine-Piccolo},
  {Jordan}, {Korn}, {Krone-Martins}, {Lanzafame}, {Lebzelter}, {L{\"o}ffler},
  {Manteiga}, {Marrese}, {Mart{\'\i}n-Fleitas}, {Moitinho}, {Mora}, {Muinonen},
  {Osinde}, {Pancino}, {Pauwels}, {Petit}, {Recio-Blanco}, {Richards},
  {Rimoldini}, {Robin}, {Sarro}, {Siopis}, {Smith}, {Sozzetti}, {S{\"u}veges},
  {Torra}, {van Reeven}, {Abbas}, {Abreu Aramburu}, {Accart}, {Aerts},
  {Altavilla}, {{\'A}lvarez}, {Alvarez}, {Alves}, {Anderson}, {Andrei},
  {Anglada Varela}, {Antiche}, {Antoja}, {Arcay}, {Astraatmadja}, {Bach},
  {Baker}, {Balaguer-N{\'u}{\~n}ez}, {Balm}, {Barache}, {Barata}, {Barbato},
  {Barblan}, {Barklem}, {Barrado}, {Barros}, {Barstow}, {Bartholom{\'e}
  Mu{\~n}oz}, {Bassilana}, {Becciani}, {Bellazzini}, {Berihuete}, {Bertone},
  {Bianchi}, {Bienaym{\'e}}, {Blanco-Cuaresma}, {Boch}, {Boeche}, {Bombrun},
  {Borrachero}, {Bossini}, {Bouquillon}, {Bourda}, {Bragaglia}, {Bramante},
  {Breddels}, {Bressan}, {Brouillet}, {Br{\"u}semeister}, {Brugaletta},
  {Bucciarelli}, {Burlacu}, {Busonero}, {Butkevich}, {Buzzi}, {Caffau},
  {Cancelliere}, {Cannizzaro}, {Cantat-Gaudin}, {Carballo}, {Carlucci},
  {Carrasco}, {Casamiquela}, {Castellani}, {Castro-Ginard}, {Charlot},
  {Chemin}, {Chiavassa}, {Cocozza}, {Costigan}, {Cowell}, {Crifo}, {Crosta},
  {Crowley}, {Cuypers}, {Dafonte}, {Damerdji}, {Dapergolas}, {David}, {David},
  {de Laverny}, {De Luise}, {De March}, {de Martino}, {de Souza}, {de Torres},
  {Debosscher}, {del Pozo}, {Delbo}, {Delgado}, {Delgado}, {Di Matteo},
  {Diakite}, {Diener}, {Distefano}, {Dolding}, {Drazinos}, {Dur{\'a}n},
  {Edvardsson}, {Enke}, {Eriksson}, {Esquej}, {Eynard Bontemps}, {Fabre},
  {Fabrizio}, {Faigler}, {Falc{\~a}o}, {Farr{\`a}s Casas}, {Federici},
  {Fedorets}, {Fernique}, {Figueras}, {Filippi}, {Findeisen}, {Fonti},
  {Fraile}, {Fraser}, {Fr{\'e}zouls}, {Gai}, {Galleti}, {Garabato},
  {Garc{\'\i}a-Sedano}, {Garofalo}, {Garralda}, {Gavel}, {Gavras}, {Gerssen},
  {Geyer}, {Giacobbe}, {Gilmore}, {Girona}, {Giuffrida}, {Glass}, {Gomes},
  {Granvik}, {Gueguen}, {Guerrier}, {Guiraud}, {Guti{\'e}rrez-S{\'a}nchez},
  {Haigron}, {Hatzidimitriou}, {Hauser}, {Haywood}, {Heiter}, {Helmi}, {Heu},
  {Hilger}, {Hobbs}, {Hofmann}, {Holland}, {Huckle}, {Hypki}, {Icardi},
  {Jan{\ss}en}, {Jevardat de Fombelle}, {Jonker}, {Juh{\'a}sz}, {Julbe},
  {Karampelas}, {Kewley}, {Klar}, {Kochoska}, {Kohley}, {Kolenberg},
  {Kontizas}, {Kontizas}, {Koposov}, {Kordopatis}, {Kostrzewa-Rutkowska},
  {Koubsky}, {Lambert}, {Lanza}, {Lasne}, {Lavigne}, {Le Fustec}, {Le
  Poncin-Lafitte}, {Lebreton}, {Leccia}, {Leclerc}, {Lecoeur-Taibi},
  {Lenhardt}, {Leroux}, {Liao}, {Licata}, {Lindstr{\o}m}, {Lister}, {Livanou},
  {Lobel}, {L{\'o}pez}, {Managau}, {Mann}, {Mantelet}, {Marchal}, {Marchant},
  {Marconi}, {Marinoni}, {Marschalk{\'o}}, {Marshall}, {Martino}, {Marton},
  {Mary}, {Massari}, {Matijevi{\v{c}}}, {Mazeh}, {McMillan}, {Messina},
  {Michalik}, {Millar}, {Molina}, {Molinaro}, {Moln{\'a}r}, {Montegriffo},
  {Mor}, {Morbidelli}, {Morel}, {Morris}, {Mulone}, {Muraveva}, {Musella},
  {Nelemans}, {Nicastro}, {Noval}, {O'Mullane}, {Ord{\'e}novic},
  {Ord{\'o}{\~n}ez-Blanco}, {Osborne}, {Pagani}, {Pagano}, {Pailler},
  {Palacin}, {Palaversa}, {Panahi}, {Pawlak}, {Piersimoni}, {Pineau}, {Plachy},
  {Plum}, {Poggio}, {Poujoulet}, {Pr{\v{s}}a}, {Pulone}, {Racero}, {Ragaini},
  {Rambaux}, {Ramos-Lerate}, {Regibo}, {Reyl{\'e}}, {Riclet}, {Ripepi}, {Riva},
  {Rivard}, {Rixon}, {Roegiers}, {Roelens}, {Romero-G{\'o}mez}, {Rowell},
  {Royer}, {Ruiz-Dern}, {Sadowski}, {Sagrist{\`a} Sell{\'e}s}, {Sahlmann},
  {Salgado}, {Salguero}, {Sanna}, {Santana-Ros}, {Sarasso}, {Savietto},
  {Schultheis}, {Sciacca}, {Segol}, {Segovia}, {S{\'e}gransan}, {Shih},
  {Siltala}, {Silva}, {Smart}, {Smith}, {Solano}, {Solitro}, {Sordo}, {Soria
  Nieto}, {Souchay}, {Spagna}, {Spoto}, {Stampa}, {Steele},
  {Steidelm{\"u}ller}, {Stephenson}, {Stoev}, {Suess}, {Surdej}, {Szabados},
  {Szegedi-Elek}, {Tapiador}, {Taris}, {Tauran}, {Taylor}, {Teixeira},
  {Terrett}, {Teyssand ier}, {Thuillot}, {Titarenko}, {Torra Clotet}, {Turon},
  {Ulla}, {Utrilla}, {Uzzi}, {Vaillant}, {Valentini}, {Valette}, {van Elteren},
  {Van Hemelryck}, {van Leeuwen}, {Vaschetto}, {Vecchiato}, {Veljanoski},
  {Viala}, {Vicente}, {Vogt}, {von Essen}, {Voss}, {Votruba}, {Voutsinas},
  {Walmsley}, {Weiler}, {Wertz}, {Wevers}, {Wyrzykowski}, {Yoldas},
  {{\v{Z}}erjal}, {Ziaeepour}, {Zorec}, {Zschocke}, {Zucker}, {Zurbach}, \&
  {Zwitter}}]{2018A&A...616A...1G}
{Gaia Collaboration}, {Brown}, A.~G.~A., {Vallenari}, A., {et~al.} 2018, \aap,
  616, A1, \dodoi{10.1051/0004-6361/201833051}

\bibitem[{{Gaia Collaboration} {et~al.}(2019{\natexlab{a}}){Gaia
  Collaboration}, Eyer, Rimoldini, Audard, Anderson, Nienartowicz, Glass,
  Marchal, Grenon, Mowlavi, Holl, Clementini, Aerts, Mazeh, Evans, Szabados,
  Brown, Vallenari, Prusti, de~Bruijne, Babusiaux, Bailer-Jones, Biermann,
  Jansen, Jordi, Klioner, Lammers, Lindegren, Luri, Mignard, Panem, Pourbaix,
  Randich, Sartoretti, Siddiqui, Soubiran, van Leeuwen, Walton, Arenou,
  Bastian, Cropper, Drimmel, Katz, Lattanzi, Bakker, Cacciari,
  Casta\$\sim${n}eda, Chaoul, Cheek, De~Angeli, Fabricius, Guerra, Masana,
  Messineo, Panuzzo, Portell, Riello, Seabroke, Tanga, Th{\'e}venin,
  Gracia-Abril, Comoretto, Garcia-Reinaldos, Teyssier, Altmann, Andrae,
  Bellas-Velidis, Benson, Berthier, Blomme, Burgess, Busso, Carry, Cellino,
  Clotet, Creevey, Davidson, De~Ridder, Delchambre, Dell’Oro, Ducourant,
  Fern{\'a}ndez-Hern{\'a}ndez, Fouesneau, Fr{\'e}mat, Galluccio,
  Garc\'{i}a-Torres, Gonz{\'a}lez-N\'{u}\$\sim${n}ez, Gonz{\'a}lez-Vidal,
  Gosset, Guy, Halbwachs, Hambly, Harrison, Hern{\'a}ndez, Hestroffer, Hodgkin,
  Hutton, Jasniewicz, Jean-Antoine-Piccolo, Jordan, Korn, Krone-Martins,
  Lanzafame, Lebzelter, L{\"o}ffler, Manteiga, Marrese, Mart\'{i}n-Fleitas,
  Moitinho, Mora, Muinonen, Osinde, Pancino, Pauwels, Petit, Recio-Blanco,
  Richards, Robin, Sarro, Siopis, Smith, Sozzetti, S{\"u}veges, Torra, van
  Reeven, Abbas, Abreu~Aramburu, Accart, Altavilla, \'{A}lvarez, Alvarez,
  Alves, Andrei, Anglada~Varela, Antiche, Antoja, Arcay, Astraatmadja, Bach,
  Baker, Balaguer-N\'{u}\$\sim${n}ez, Balm, Barache, Barata, Barbato, Barblan,
  Barklem, Barrado, Barros, Barstow, Bartholom{\'e}~Mu\$\sim${n}oz, Bassilana,
  Becciani, Bellazzini, Berihuete, Bertone, Bianchi, Bienaym{\'e},
  Blanco-Cuaresma, Boch, Boeche, Bombrun, Borrachero, Bossini, Bouquillon,
  Bourda, Bragaglia, Bramante, Breddels, Bressan, Brouillet, Br{\"u}semeister,
  Brugaletta, Bucciarelli, Burlacu, Busonero, Butkevich, Buzzi, Caffau,
  Cancelliere, Cannizzaro, Cantat-Gaudin, Carballo, Carlucci, Carrasco,
  Casamiquela, Castellani, Castro-Ginard, Charlot, Chemin, Chiavassa, Cocozza,
  Costigan, Cowell, Crifo, Crosta, Crowley, Cuypers, Dafonte, Damerdji,
  Dapergolas, David, David, de~Laverny, De~Luise, De~March, de~Martino,
  de~Souza, de~Torres, Debosscher, del Pozo, Delbo, Delgado, Delgado, Diakite,
  Diener, Distefano, Dolding, Drazinos, Dur{\'a}n, Edvardsson, Enke, Eriksson,
  Esquej, Eynard~Bontemps, Fabre, Fabrizio, Faigler, Falc\$\sim${a}o,
  Farr{\`a}s~Casas, Federici, Fedorets, Fernique, Figueras, Filippi, Findeisen,
  Fonti, Fraile, Fraser, Fr{\'e}zouls, Gai, Galleti, Garabato,
  Garc\'{i}a-Sedano, Garofalo, Garralda, Gavel, Gavras, Gerssen, Geyer,
  Giacobbe, Gilmore, Girona, Giuffrida, Gomes, Granvik, Gueguen, Guerrier,
  Guiraud, Guti{\'e}rrez-S{\'a}nchez, Haigron, Hatzidimitriou, Hauser, Haywood,
  Heiter, Helmi, Heu, Hilger, Hobbs, Hofmann, Holland, Huckle, Hypki, Icardi,
  Janßen, Jevardat~de Fombelle, Jonker, Juh{\'a}sz, Julbe, Karampelas, Kewley,
  Klar, Kochoska, Kohley, Kolenberg, Kontizas, Kontizas, Koposov, Kordopatis,
  Kostrzewa-Rutkowska, Koubsky, Lambert, Lanza, Lasne, Lavigne, Le~Fustec,
  Le~Poncin-Lafitte, Lebreton, Leccia, Leclerc, Lecoeur-Taibi, Lenhardt,
  Leroux, Liao, Licata, Lindstr{\o}m, Lister, Livanou, Lobel, L{\'o}pez,
  Lorenz, Managau, Mann, Mantelet, Marchant, Marconi, Marinoni, Marschalk{\'o},
  Marshall, Martino, Marton, Mary, Massari, Matijevič, McMillan, Messina,
  Michalik, Millar, Molina, Molinaro, Moln{\'a}r, Montegriffo, Mor, Morbidelli,
  Morel, Morgenthaler, Morris, Mulone, Muraveva, Musella, Nelemans, Nicastro,
  Noval, O’Mullane, Ord{\'e}novic, Ord{\'o}\$\sim${n}ez-Blanco, Osborne,
  Pagani, Pagano, Pailler, Palacin, Palaversa, Panahi, Pawlak, Piersimoni,
  Pineau, Plachy, Plum, Poggio, Poujoulet, Pr\v{s}a, Pulone, Racero, Ragaini,
  Rambaux, Ramos-Lerate, Regibo, Reyl{\'e}, Riclet, Ripepi, Riva, Rivard,
  Rixon, Roegiers, Roelens, Romero-G{\'o}mez, Rowell, Royer, Ruiz-Dern,
  Sadowski, Sagrist{\`a}~Sell{\'e}s, Sahlmann, Salgado, Salguero, Sanna,
  Santana-Ros, Sarasso, Savietto, Schultheis, Sciacca, Segol, Segovia,
  S{\'e}gransan, Shih, Siltala, Silva, Smart, Smith, Solano, Solitro, Sordo,
  Soria~Nieto, Souchay, Spagna, Spoto, Stampa, Steele, Steidelm{\"u}ller,
  Stephenson, Stoev, Suess, Surdej, Szegedi-Elek, Tapiador, Taris, Tauran,
  Taylor, Teixeira, Terrett, Teyssandier, Thuillot, Titarenko, Torra~Clotet,
  Turon, Ulla, Utrilla, Uzzi, Vaillant, Valentini, Valette, van Elteren,
  Van~Hemelryck, van Leeuwen, Vaschetto, Vecchiato, Veljanoski, Viala, Vicente,
  Vogt, von Essen, Voss, Votruba, Voutsinas, Walmsley, Weiler, Wertz, Wevers,
  Wyrzykowski, Yoldas, \v{Z}erjal, Ziaeepour, Zorec, Zschocke, Zucker, Zurbach,
  \& Zwitter}]{gaia_collaboration_gaia_2019}
{Gaia Collaboration}, Eyer, L., Rimoldini, L., {et~al.} 2019{\natexlab{a}},
  \aap, 623, A110, \dodoi{10.1051/0004-6361/201833304}

\bibitem[{{Gaia Collaboration} {et~al.}(2019{\natexlab{b}}){Gaia
  Collaboration}, {Eyer}, {Rimoldini}, {Audard}, {Anderson}, {Nienartowicz},
  {Glass}, {Marchal}, {Grenon}, {Mowlavi}, {Holl}, {Clementini}, {Aerts},
  {Mazeh}, {Evans}, {Szabados}, {Brown}, {Vallenari}, {Prusti}, {de Bruijne},
  {Babusiaux}, {Bailer-Jones}, {Biermann}, {Jansen}, {Jordi}, {Klioner},
  {Lammers}, {Lindegren}, {Luri}, {Mignard}, {Panem}, {Pourbaix}, {Randich},
  {Sartoretti}, {Siddiqui}, {Soubiran}, {van Leeuwen}, {Walton}, {Arenou},
  {Bastian}, {Cropper}, {Drimmel}, {Katz}, {Lattanzi}, {Bakker}, {Cacciari},
  {Casta{\~n}eda}, {Chaoul}, {Cheek}, {De Angeli}, {Fabricius}, {Guerra},
  {Masana}, {Messineo}, {Panuzzo}, {Portell}, {Riello}, {Seabroke}, {Tanga},
  {Th{\'e}venin}, {Gracia-Abril}, {Comoretto}, {Garcia-Reinaldos}, {Teyssier},
  {Altmann}, {Andrae}, {Bellas-Velidis}, {Benson}, {Berthier}, {Blomme},
  {Burgess}, {Busso}, {Carry}, {Cellino}, {Clotet}, {Creevey}, {Davidson}, {De
  Ridder}, {Delchambre}, {Dell'Oro}, {Ducourant},
  {Fern{\'a}ndez-Hern{\'a}ndez}, {Fouesneau}, {Fr{\'e}mat}, {Galluccio},
  {Garc{\'\i}a-Torres}, {Gonz{\'a}lez-N{\'u}{\~n}ez}, {Gonz{\'a}lez-Vidal},
  {Gosset}, {Guy}, {Halbwachs}, {Hambly}, {Harrison}, {Hern{\'a}ndez},
  {Hestroffer}, {Hodgkin}, {Hutton}, {Jasniewicz}, {Jean-Antoine-Piccolo},
  {Jordan}, {Korn}, {Krone-Martins}, {Lanzafame}, {Lebzelter}, {L{\"o}ffler},
  {Manteiga}, {Marrese}, {Mart{\'\i}n-Fleitas}, {Moitinho}, {Mora}, {Muinonen},
  {Osinde}, {Pancino}, {Pauwels}, {Petit}, {Recio-Blanco}, {Richards}, {Robin},
  {Sarro}, {Siopis}, {Smith}, {Sozzetti}, {S{\"u}veges}, {Torra}, {van Reeven},
  {Abbas}, {Abreu Aramburu}, {Accart}, {Altavilla}, {{\'A}lvarez}, {Alvarez},
  {Alves}, {Andrei}, {Anglada Varela}, {Antiche}, {Antoja}, {Arcay},
  {Astraatmadja}, {Bach}, {Baker}, {Balaguer-N{\'u}{\~n}ez}, {Balm}, {Barache},
  {Barata}, {Barbato}, {Barblan}, {Barklem}, {Barrado}, {Barros}, {Barstow},
  {Bartholom{\'e} Mu{\~n}oz}, {Bassilana}, {Becciani}, {Bellazzini},
  {Berihuete}, {Bertone}, {Bianchi}, {Bienaym{\'e}}, {Blanco-Cuaresma}, {Boch},
  {Boeche}, {Bombrun}, {Borrachero}, {Bossini}, {Bouquillon}, {Bourda},
  {Bragaglia}, {Bramante}, {Breddels}, {Bressan}, {Brouillet},
  {Br{\"u}semeister}, {Brugaletta}, {Bucciarelli}, {Burlacu}, {Busonero},
  {Butkevich}, {Buzzi}, {Caffau}, {Cancelliere}, {Cannizzaro}, {Cantat-Gaudin},
  {Carballo}, {Carlucci}, {Carrasco}, {Casamiquela}, {Castellani},
  {Castro-Ginard}, {Charlot}, {Chemin}, {Chiavassa}, {Cocozza}, {Costigan},
  {Cowell}, {Crifo}, {Crosta}, {Crowley}, {Cuypers}, {Dafonte}, {Damerdji},
  {Dapergolas}, {David}, {David}, {de Laverny}, {De Luise}, {De March}, {de
  Martino}, {de Souza}, {de Torres}, {Debosscher}, {del Pozo}, {Delbo},
  {Delgado}, {Delgado}, {Diakite}, {Diener}, {Distefano}, {Dolding},
  {Drazinos}, {Dur{\'a}n}, {Edvardsson}, {Enke}, {Eriksson}, {Esquej}, {Eynard
  Bontemps}, {Fabre}, {Fabrizio}, {Faigler}, {Falc{\~a}o}, {Farr{\`a}s Casas},
  {Federici}, {Fedorets}, {Fernique}, {Figueras}, {Filippi}, {Findeisen},
  {Fonti}, {Fraile}, {Fraser}, {Fr{\'e}zouls}, {Gai}, {Galleti}, {Garabato},
  {Garc{\'\i}a-Sedano}, {Garofalo}, {Garralda}, {Gavel}, {Gavras}, {Gerssen},
  {Geyer}, {Giacobbe}, {Gilmore}, {Girona}, {Giuffrida}, {Gomes}, {Granvik},
  {Gueguen}, {Guerrier}, {Guiraud}, {Guti{\'e}rrez-S{\'a}nchez}, {Haigron},
  {Hatzidimitriou}, {Hauser}, {Haywood}, {Heiter}, {Helmi}, {Heu}, {Hilger},
  {Hobbs}, {Hofmann}, {Holland}, {Huckle}, {Hypki}, {Icardi}, {Jan{\ss}en},
  {Jevardat de Fombelle}, {Jonker}, {Juh{\'a}sz}, {Julbe}, {Karampelas},
  {Kewley}, {Klar}, {Kochoska}, {Kohley}, {Kolenberg}, {Kontizas}, {Kontizas},
  {Koposov}, {Kordopatis}, {Kostrzewa-Rutkowska}, {Koubsky}, {Lambert},
  {Lanza}, {Lasne}, {Lavigne}, {Le Fustec}, {Le Poncin-Lafitte}, {Lebreton},
  {Leccia}, {Leclerc}, {Lecoeur-Taibi}, {Lenhardt}, {Leroux}, {Liao}, {Licata},
  {Lindstr{\o}m}, {Lister}, {Livanou}, {Lobel}, {L{\'o}pez}, {Lorenz},
  {Managau}, {Mann}, {Mantelet}, {Marchant}, {Marconi}, {Marinoni},
  {Marschalk{\'o}}, {Marshall}, {Martino}, {Marton}, {Mary}, {Massari},
  {Matijevi{\v{c}}}, {McMillan}, {Messina}, {Michalik}, {Millar}, {Molina},
  {Molinaro}, {Moln{\'a}r}, {Montegriffo}, {Mor}, {Morbidelli}, {Morel},
  {Morgenthaler}, {Morris}, {Mulone}, {Muraveva}, {Musella}, {Nelemans},
  {Nicastro}, {Noval}, {O'Mullane}, {Ord{\'e}novic}, {Ord{\'o}{\~n}ez-Blanco},
  {Osborne}, {Pagani}, {Pagano}, {Pailler}, {Palacin}, {Palaversa}, {Panahi},
  {Pawlak}, {Piersimoni}, {Pineau}, {Plachy}, {Plum}, {Poggio}, {Poujoulet},
  {Pr{\v{s}}a}, {Pulone}, {Racero}, {Ragaini}, {Rambaux}, {Ramos-Lerate},
  {Regibo}, {Reyl{\'e}}, {Riclet}, {Ripepi}, {Riva}, {Rivard}, {Rixon},
  {Roegiers}, {Roelens}, {Romero-G{\'o}mez}, {Rowell}, {Royer}, {Ruiz-Dern},
  {Sadowski}, {Sagrist{\`a} Sell{\'e}s}, {Sahlmann}, {Salgado}, {Salguero},
  {Sanna}, {Santana-Ros}, {Sarasso}, {Savietto}, {Schultheis}, {Sciacca},
  {Segol}, {Segovia}, {S{\'e}gransan}, {Shih}, {Siltala}, {Silva}, {Smart},
  {Smith}, {Solano}, {Solitro}, {Sordo}, {Soria Nieto}, {Souchay}, {Spagna},
  {Spoto}, {Stampa}, {Steele}, {Steidelm{\"u}ller}, {Stephenson}, {Stoev},
  {Suess}, {Surdej}, {Szegedi-Elek}, {Tapiador}, {Taris}, {Tauran}, {Taylor},
  {Teixeira}, {Terrett}, {Teyssandier}, {Thuillot}, {Titarenko}, {Torra
  Clotet}, {Turon}, {Ulla}, {Utrilla}, {Uzzi}, {Vaillant}, {Valentini},
  {Valette}, {van Elteren}, {Van Hemelryck}, {van Leeuwen}, {Vaschetto},
  {Vecchiato}, {Veljanoski}, {Viala}, {Vicente}, {Vogt}, {von Essen}, {Voss},
  {Votruba}, {Voutsinas}, {Walmsley}, {Weiler}, {Wertz}, {Wevers},
  {Wyrzykowski}, {Yoldas}, {{\v{Z}}erjal}, {Ziaeepour}, {Zorec}, {Zschocke},
  {Zucker}, {Zurbach}, \& {Zwitter}}]{2019A&A...623A.110G}
{Gaia Collaboration}, {Eyer}, L., {Rimoldini}, L., {et~al.} 2019{\natexlab{b}},
  \aap, 623, A110, \dodoi{10.1051/0004-6361/201833304}

\bibitem[{{Goldstein} {et~al.}(2015){Goldstein}, {D'Andrea}, {Fischer},
  {Foley}, {Gupta}, {Kessler}, {Kim}, {Nichol}, {Nugent}, {Papadopoulos},
  {Sako}, {Smith}, {Sullivan}, {Thomas}, {Wester}, {Wolf}, {Abdalla},
  {Banerji}, {Benoit-L{\'e}vy}, {Bertin}, {Brooks}, {Carnero Rosell},
  {Castander}, {da Costa}, {Covarrubias}, {DePoy}, {Desai}, {Diehl}, {Doel},
  {Eifler}, {Fausti Neto}, {Finley}, {Flaugher}, {Fosalba}, {Frieman},
  {Gerdes}, {Gruen}, {Gruendl}, {James}, {Kuehn}, {Kuropatkin}, {Lahav}, {Li},
  {Maia}, {Makler}, {March}, {Marshall}, {Martini}, {Merritt}, {Miquel},
  {Nord}, {Ogando}, {Plazas}, {Romer}, {Roodman}, {Sanchez}, {Scarpine},
  {Schubnell}, {Sevilla-Noarbe}, {Smith}, {Soares-Santos}, {Sobreira},
  {Suchyta}, {Swanson}, {Tarle}, {Thaler}, \& {Walker}}]{2015AJ....150...82G}
{Goldstein}, D.~A., {D'Andrea}, C.~B., {Fischer}, J.~A., {et~al.} 2015, \aj,
  150, 82, \dodoi{10.1088/0004-6256/150/3/82}

\bibitem[{Goodfellow {et~al.}(2016)Goodfellow, Bengio, \&
  Courville}]{Goodfellow-et-al-2016}
Goodfellow, I., Bengio, Y., \& Courville, A. 2016, Deep Learning (MIT Press)

\bibitem[{{Goodfellow} {et~al.}(2014){Goodfellow}, {Pouget-Abadie}, {Mirza},
  {Xu}, {Warde-Farley}, {Ozair}, {Courville}, \&
  {Bengio}}]{2014arXiv1406.2661G}
{Goodfellow}, I.~J., {Pouget-Abadie}, J., {Mirza}, M., {et~al.} 2014, arXiv
  e-prints, arXiv:1406.2661.
\newblock \doarXiv{1406.2661}

\bibitem[{{Guillochon} {et~al.}(2018){Guillochon}, {Nicholl}, {Villar},
  {Mockler}, {Narayan}, {Mandel}, {Berger}, \&
  {Williams}}]{2018ApJS..236....6G}
{Guillochon}, J., {Nicholl}, M., {Villar}, V.~A., {et~al.} 2018, \apjs, 236, 6,
  \dodoi{10.3847/1538-4365/aab761}

\bibitem[{{Guo} {et~al.}(2019){Guo}, {Duan}, {Wang}, {Yao}, {Yin}, {Xin}, {Li},
  {Qian}, {Wang}, {Pan}, \& {Zhang}}]{2019MNRAS.490.5424G}
{Guo}, P., {Duan}, F., {Wang}, P., {et~al.} 2019, \mnras, 490, 5424,
  \dodoi{10.1093/mnras/stz2975}

\bibitem[{He {et~al.}(2016)He, Zhang, Ren, \& Sun}]{He_2016_CVPR}
He, K., Zhang, X., Ren, S., \& Sun, J. 2016, in The IEEE Conference on Computer
  Vision and Pattern Recognition (CVPR)

\bibitem[{{Hinton} \& {Salakhutdinov}(2006)}]{2006Sci...313..504H}
{Hinton}, G.~E., \& {Salakhutdinov}, R.~R. 2006, Science, 313, 504,
  \dodoi{10.1126/science.1127647}

\bibitem[{Hochreiter \& Schmidhuber(1997)}]{lstm}
Hochreiter, S., \& Schmidhuber, J. 1997, Neural computation, 9, 1735,
  \dodoi{10.1162/neco.1997.9.8.1735}

\bibitem[{Hunter(2007)}]{Hunter:2007}
Hunter, J.~D. 2007, Computing in Science \& Engineering, 9, 90,
  \dodoi{10.1109/MCSE.2007.55}

\bibitem[{{Ichinohe} \& {Yamada}(2019)}]{2019MNRAS.487.2874I}
{Ichinohe}, Y., \& {Yamada}, S. 2019, \mnras, 487, 2874,
  \dodoi{10.1093/mnras/stz1528}

\bibitem[{{Ivezi{\'c}} {et~al.}(2019){Ivezi{\'c}}, {Kahn}, {Tyson}, {Abel},
  {Acosta}, {Allsman}, {Alonso}, {AlSayyad}, {Anderson}, {Andrew}, {Angel},
  {Angeli}, {Ansari}, {Antilogus}, {Araujo}, {Armstrong}, {Arndt}, {Astier},
  {Aubourg}, {Auza}, {Axelrod}, {Bard}, {Barr}, {Barrau}, {Bartlett}, {Bauer},
  {Bauman}, {Baumont}, {Bechtol}, {Bechtol}, {Becker}, {Becla}, {Beldica},
  {Bellavia}, {Bianco}, {Biswas}, {Blanc}, {Blazek}, {Bland ford}, {Bloom},
  {Bogart}, {Bond}, {Booth}, {Borgland}, {Borne}, {Bosch}, {Boutigny},
  {Brackett}, {Bradshaw}, {Brand t}, {Brown}, {Bullock}, {Burchat}, {Burke},
  {Cagnoli}, {Calabrese}, {Callahan}, {Callen}, {Carlin}, {Carlson}, {Chand
  rasekharan}, {Charles-Emerson}, {Chesley}, {Cheu}, {Chiang}, {Chiang},
  {Chirino}, {Chow}, {Ciardi}, {Claver}, {Cohen-Tanugi}, {Cockrum}, {Coles},
  {Connolly}, {Cook}, {Cooray}, {Covey}, {Cribbs}, {Cui}, {Cutri}, {Daly},
  {Daniel}, {Daruich}, {Daubard}, {Daues}, {Dawson}, {Delgado}, {Dellapenna},
  {de Peyster}, {de Val-Borro}, {Digel}, {Doherty}, {Dubois},
  {Dubois-Felsmann}, {Durech}, {Economou}, {Eifler}, {Eracleous}, {Emmons},
  {Fausti Neto}, {Ferguson}, {Figueroa}, {Fisher-Levine}, {Focke}, {Foss},
  {Frank}, {Freemon}, {Gangler}, {Gawiser}, {Geary}, {Gee}, {Geha}, {Gessner},
  {Gibson}, {Gilmore}, {Glanzman}, {Glick}, {Goldina}, {Goldstein}, {Goodenow},
  {Graham}, {Gressler}, {Gris}, {Guy}, {Guyonnet}, {Haller}, {Harris},
  {Hascall}, {Haupt}, {Hernand ez}, {Herrmann}, {Hileman}, {Hoblitt},
  {Hodgson}, {Hogan}, {Howard}, {Huang}, {Huffer}, {Ingraham}, {Innes},
  {Jacoby}, {Jain}, {Jammes}, {Jee}, {Jenness}, {Jernigan}, {Jevremovi{\'c}},
  {Johns}, {Johnson}, {Johnson}, {Jones}, {Juramy-Gilles}, {Juri{\'c}},
  {Kalirai}, {Kallivayalil}, {Kalmbach}, {Kantor}, {Karst}, {Kasliwal},
  {Kelly}, {Kessler}, {Kinnison}, {Kirkby}, {Knox}, {Kotov}, {Krabbendam},
  {Krughoff}, {Kub{\'a}nek}, {Kuczewski}, {Kulkarni}, {Ku}, {Kurita}, {Lage},
  {Lambert}, {Lange}, {Langton}, {Le Guillou}, {Levine}, {Liang}, {Lim},
  {Lintott}, {Long}, {Lopez}, {Lotz}, {Lupton}, {Lust}, {MacArthur}, {Mahabal},
  {Mand elbaum}, {Markiewicz}, {Marsh}, {Marshall}, {Marshall}, {May},
  {McKercher}, {McQueen}, {Meyers}, {Migliore}, {Miller}, {Mills}, {Miraval},
  {Moeyens}, {Moolekamp}, {Monet}, {Moniez}, {Monkewitz}, {Montgomery},
  {Morrison}, {Mueller}, {Muller}, {Mu{\~n}oz Arancibia}, {Neill}, {Newbry},
  {Nief}, {Nomerotski}, {Nordby}, {O'Connor}, {Oliver}, {Olivier}, {Olsen},
  {O'Mullane}, {Ortiz}, {Osier}, {Owen}, {Pain}, {Palecek}, {Parejko},
  {Parsons}, {Pease}, {Peterson}, {Peterson}, {Petravick}, {Libby Petrick},
  {Petry}, {Pierfederici}, {Pietrowicz}, {Pike}, {Pinto}, {Plante}, {Plate},
  {Plutchak}, {Price}, {Prouza}, {Radeka}, {Rajagopal}, {Rasmussen},
  {Regnault}, {Reil}, {Reiss}, {Reuter}, {Ridgway}, {Riot}, {Ritz}, {Robinson},
  {Roby}, {Roodman}, {Rosing}, {Roucelle}, {Rumore}, {Russo}, {Saha},
  {Sassolas}, {Schalk}, {Schellart}, {Schindler}, {Schmidt}, {Schneider},
  {Schneider}, {Schoening}, {Schumacher}, {Schwamb}, {Sebag}, {Selvy},
  {Sembroski}, {Seppala}, {Serio}, {Serrano}, {Shaw}, {Shipsey}, {Sick},
  {Silvestri}, {Slater}, {Smith}, {Smith}, {Sobhani}, {Soldahl},
  {Storrie-Lombardi}, {Stover}, {Strauss}, {Street}, {Stubbs}, {Sullivan},
  {Sweeney}, {Swinbank}, {Szalay}, {Takacs}, {Tether}, {Thaler}, {Thayer},
  {Thomas}, {Thornton}, {Thukral}, {Tice}, {Trilling}, {Turri}, {Van Berg},
  {Vanden Berk}, {Vetter}, {Virieux}, {Vucina}, {Wahl}, {Walkowicz}, {Walsh},
  {Walter}, {Wang}, {Wang}, {Warner}, {Wiecha}, {Willman}, {Winters},
  {Wittman}, {Wolff}, {Wood-Vasey}, {Wu}, {Xin}, {Yoachim}, \&
  {Zhan}}]{2019ApJ...873..111I}
{Ivezi{\'c}}, {\v{Z}}., {Kahn}, S.~M., {Tyson}, J.~A., {et~al.} 2019, \apj,
  873, 111, \dodoi{10.3847/1538-4357/ab042c}

\bibitem[{{Jamal} \& {Bloom}(2020)}]{2020arXiv200308618J}
{Jamal}, S., \& {Bloom}, J.~S. 2020, arXiv e-prints, arXiv:2003.08618.
\newblock \doarXiv{2003.08618}

\bibitem[{{Jimenez Rezende} \& {Mohamed}(2015)}]{2015arXiv150505770J}
{Jimenez Rezende}, D., \& {Mohamed}, S. 2015, arXiv e-prints, arXiv:1505.05770.
\newblock \doarXiv{1505.05770}

\bibitem[{{Karras} {et~al.}(2019){Karras}, {Laine}, {Aittala}, {Hellsten},
  {Lehtinen}, \& {Aila}}]{2019arXiv191204958K}
{Karras}, T., {Laine}, S., {Aittala}, M., {et~al.} 2019, arXiv e-prints,
  arXiv:1912.04958.
\newblock \doarXiv{1912.04958}

\bibitem[{{Kessler} {et~al.}(2009){Kessler}, {Bernstein}, {Cinabro}, {Dilday},
  {Frieman}, {Jha}, {Kuhlmann}, {Miknaitis}, {Sako}, {Taylor}, \&
  {Vanderplas}}]{2009PASP..121.1028K}
{Kessler}, R., {Bernstein}, J.~P., {Cinabro}, D., {et~al.} 2009, \pasp, 121,
  1028, \dodoi{10.1086/605984}

\bibitem[{{Kessler} {et~al.}(2019){Kessler}, {Narayan}, {Avelino}, {Bachelet},
  {Biswas}, {Brown}, {Chernoff}, {Connolly}, {Dai}, {Daniel}, {Di Stefano},
  {Drout}, {Galbany}, {Gonz{\'a}lez-Gait{\'a}n}, {Graham}, {Hlo{\v{z}}ek},
  {Ishida}, {Guillochon}, {Jha}, {Jones}, {Mand el}, {Muthukrishna}, {O'Grady},
  {Peters}, {Pierel}, {Ponder}, {Pr{\v{s}}a}, {Rodney}, {Villar}, {LSST Dark
  Energy Science Collaboration}, \& {Transient and Variable Stars Science
  Collaboration}}]{2019PASP..131i4501K}
{Kessler}, R., {Narayan}, G., {Avelino}, A., {et~al.} 2019, \pasp, 131, 094501,
  \dodoi{10.1088/1538-3873/ab26f1}

\bibitem[{{Kingma} \& {Ba}(2014)}]{2014arXiv1412.6980K}
{Kingma}, D.~P., \& {Ba}, J. 2014, arXiv e-prints, arXiv:1412.6980.
\newblock \doarXiv{1412.6980}

\bibitem[{{Kingma} \& {Welling}(2013)}]{2013arXiv1312.6114K}
{Kingma}, D.~P., \& {Welling}, M. 2013, arXiv e-prints, arXiv:1312.6114.
\newblock \doarXiv{1312.6114}

\bibitem[{Kluyver {et~al.}(2016)Kluyver, Ragan-Kelley, Perez, Granger,
  Bussonnier, Frederic, Kelley, Hamrick, Grout, Corlay, Ivanov, Avila, Abdalla,
  Willing, \& [Unknown}]{jupyter}
Kluyver, T., Ragan-Kelley, B., Perez, F., {et~al.} 2016

\bibitem[{Kullback \& Leibler(1951)}]{kullback1951}
Kullback, S., \& Leibler, R.~A. 1951, Ann. Math. Statist., 22, 79,
  \dodoi{10.1214/aoms/1177729694}

\bibitem[{{Kumar} {et~al.}(2014){Kumar}, {MacDonald}, {Brown}, {Pfeiffer},
  {Cannon}, {Boyle}, {Kidder}, {Mrou{\'e}}, {Scheel}, {Szil{\'a}gyi}, \&
  {Zengino{\v{g}}lu}}]{2014PhRvD..89d2002K}
{Kumar}, P., {MacDonald}, I., {Brown}, D.~A., {et~al.} 2014, \prd, 89, 042002,
  \dodoi{10.1103/PhysRevD.89.042002}

\bibitem[{{Lample} {et~al.}(2017){Lample}, {Zeghidour}, {Usunier}, {Bordes},
  {Denoyer}, \& {Ranzato}}]{2017arXiv170600409L}
{Lample}, G., {Zeghidour}, N., {Usunier}, N., {et~al.} 2017, arXiv e-prints,
  arXiv:1706.00409.
\newblock \doarXiv{1706.00409}

\bibitem[{Lochner {et~al.}(2016)Lochner, McEwen, Peiris, Lahav, \&
  Winter}]{Lochner16}
Lochner, M., McEwen, J.~D., Peiris, H.~V., Lahav, O., \& Winter, M.~K. 2016,
  The Astrophysical Journal Supplement Series, 225, 31.
\newblock \url{http://stacks.iop.org/0067-0049/225/i=2/a=31}

\bibitem[{{Mahabal} {et~al.}(2019){Mahabal}, {Rebbapragada}, {Walters},
  {Masci}, {Blagorodnova}, {van Roestel}, {Ye}, {Biswas}, {Burdge}, {Chang},
  {Duev}, {Golkhou}, {Miller}, {Nordin}, {Ward}, {Adams}, {Bellm}, {Branton},
  {Bue}, {Cannella}, {Connolly}, {Dekany}, {Feindt}, {Hung}, {Fortson},
  {Frederick}, {Fremling}, {Gezari}, {Graham}, {Groom}, {Kasliwal}, {Kulkarni},
  {Kupfer}, {Lin}, {Lintott}, {Lunnan}, {Parejko}, {Prince}, {Riddle},
  {Rusholme}, {Saunders}, {Sedaghat}, {Shupe}, {Singer}, {Soumagnac}, {Szkody},
  {Tachibana}, {Tirumala}, {van Velzen}, \& {Wright}}]{2019PASP..131c8002M}
{Mahabal}, A., {Rebbapragada}, U., {Walters}, R., {et~al.} 2019, \pasp, 131,
  038002, \dodoi{10.1088/1538-3873/aaf3fa}

\bibitem[{{Mart{\'\i}nez-Palomera} {et~al.}(2018){Mart{\'\i}nez-Palomera},
  {F{\"o}rster}, {Protopapas}, {Maureira}, {Lira}, {Cabrera-Vives}, {Huijse},
  {Galbany}, {de Jaeger}, {Gonz{\'a}lez- Gait{\'a}n}, {Medina}, {Pignata}, {San
  Mart{\'\i}n}, {Hamuy}, \& {Mu{\~n}oz}}]{2018AJ....156..186M}
{Mart{\'\i}nez-Palomera}, J., {F{\"o}rster}, F., {Protopapas}, P., {et~al.}
  2018, \aj, 156, 186, \dodoi{10.3847/1538-3881/aadfd8}

\bibitem[{{Mustafa} {et~al.}(2019){Mustafa}, {Bard}, {Bhimji}, {Luki{\'c}},
  {Al-Rfou}, \& {Kratochvil}}]{2019ComAC...6....1M}
{Mustafa}, M., {Bard}, D., {Bhimji}, W., {et~al.} 2019, Computational
  Astrophysics and Cosmology, 6, 1, \dodoi{10.1186/s40668-019-0029-9}

\bibitem[{{Muthukrishna} {et~al.}(2019){Muthukrishna}, {Narayan}, {Mandel},
  {Biswas}, \& {Hlo{\v{z}}ek}}]{Muthukrishna+2019}
{Muthukrishna}, D., {Narayan}, G., {Mandel}, K.~S., {Biswas}, R., \&
  {Hlo{\v{z}}ek}, R. 2019, \pasp, 131, 118002, \dodoi{10.1088/1538-3873/ab1609}

\bibitem[{{Naul} {et~al.}(2018){Naul}, {Bloom}, {P{\'e}rez}, \& {van der
  Walt}}]{2018NatAs...2..151N}
{Naul}, B., {Bloom}, J.~S., {P{\'e}rez}, F., \& {van der Walt}, S. 2018, Nature
  Astronomy, 2, 151, \dodoi{10.1038/s41550-017-0321-z}

\bibitem[{{Nun} {et~al.}(2016){Nun}, {Protopapas}, {Sim}, \& {Chen}}]{Nun16}
{Nun}, I., {Protopapas}, P., {Sim}, B., \& {Chen}, W. 2016, \aj, 152, 71,
  \dodoi{10.3847/0004-6256/152/3/71}

\bibitem[{{Pascanu} {et~al.}(2012){Pascanu}, {Mikolov}, \&
  {Bengio}}]{2012arXiv1211.5063P}
{Pascanu}, R., {Mikolov}, T., \& {Bengio}, Y. 2012, arXiv e-prints,
  arXiv:1211.5063.
\newblock \doarXiv{1211.5063}

\bibitem[{Paszke {et~al.}(2017)Paszke, Gross, Chintala, Chanan, Yang, DeVito,
  Lin, Desmaison, Antiga, \& Lerer}]{paszke2017automatic}
Paszke, A., Gross, S., Chintala, S., {et~al.} 2017, in NIPS-W

\bibitem[{Pedregosa {et~al.}(2011)Pedregosa, Varoquaux, Gramfort, Michel,
  Thirion, Grisel, Blondel, Prettenhofer, Weiss, Dubourg, Vanderplas, Passos,
  Cournapeau, Brucher, Perrot, \& Duchesnay}]{scikit-learn}
Pedregosa, F., Varoquaux, G., Gramfort, A., {et~al.} 2011, Journal of Machine
  Learning Research, 12, 2825

\bibitem[{{Pichara} \& {Protopapas}(2013)}]{Pichara13}
{Pichara}, K., \& {Protopapas}, P. 2013, \apj, 777, 83,
  \dodoi{10.1088/0004-637X/777/2/83}

\bibitem[{{Pichara} {et~al.}(2016){Pichara}, {Protopapas}, \&
  {Le{\'o}n}}]{Pichara16}
{Pichara}, K., {Protopapas}, P., \& {Le{\'o}n}, D. 2016, \apj, 819, 18,
  \dodoi{10.3847/0004-637X/819/1/18}

\bibitem[{{Pietrukowicz} {et~al.}(2015){Pietrukowicz}, {Koz{\l}owski},
  {Skowron}, {Soszy{\'n}ski}, {Udalski}, {Poleski}, {Wyrzykowski},
  {Szyma{\'n}ski}, {Pietrzy{\'n}ski}, {Ulaczyk}, {Mr{\'o}z}, {Skowron}, \&
  {Kubiak}}]{2015ApJ...811..113P}
{Pietrukowicz}, P., {Koz{\l}owski}, S., {Skowron}, J., {et~al.} 2015, \apj,
  811, 113, \dodoi{10.1088/0004-637X/811/2/113}

\bibitem[{{Pr{\v{s}}a} {et~al.}(2016){Pr{\v{s}}a}, {Conroy}, {Horvat}, {Pablo},
  {Kochoska}, {Bloemen}, {Giammarco}, {Hambleton}, \&
  {Degroote}}]{2016ApJS..227...29P}
{Pr{\v{s}}a}, A., {Conroy}, K.~E., {Horvat}, M., {et~al.} 2016, \apjs, 227, 29,
  \dodoi{10.3847/1538-4365/227/2/29}

\bibitem[{{Rajeswar} {et~al.}(2017){Rajeswar}, {Subramanian}, {Dutil}, {Pal},
  \& {Courville}}]{2017arXiv170510929R}
{Rajeswar}, S., {Subramanian}, S., {Dutil}, F., {Pal}, C., \& {Courville}, A.
  2017, arXiv e-prints, arXiv:1705.10929.
\newblock \doarXiv{1705.10929}

\bibitem[{{Richards} {et~al.}(2012){Richards}, {Starr}, {Miller}, {Bloom},
  {Butler}, {Brink}, \& {Crellin-Quick}}]{Richards12}
{Richards}, J.~W., {Starr}, D.~L., {Miller}, A.~A., {et~al.} 2012, The
  Astrophysical Journal Supplement Series, 203, 32,
  \dodoi{10.1088/0067-0049/203/2/32}

\bibitem[{{Richards} {et~al.}(2011){Richards}, {Starr}, {Butler}, {Bloom},
  {Brewer}, {Crellin-Quick}, {Higgins}, {Kennedy}, \&
  {Rischard}}]{2011ApJ...733...10R}
{Richards}, J.~W., {Starr}, D.~L., {Butler}, N.~R., {et~al.} 2011, \apj, 733,
  10, \dodoi{10.1088/0004-637X/733/1/10}

\bibitem[{{Rimoldini} {et~al.}(2019){Rimoldini}, {Holl}, {Audard}, {Mowlavi},
  {Nienartowicz}, {Evans}, {Guy}, {Lecoeur-Ta{\"\i}bi}, {Jevardat de Fombelle},
  {Marchal}, {Roelens}, {De Ridder}, {Sarro}, {Regibo}, {Lopez}, {Clementini},
  {Ripepi}, {Molinaro}, {Garofalo}, {Moln{\'a}r}, {Plachy}, {Juh{\'a}sz},
  {Szabados}, {Lebzelter}, {Teyssier}, \& {Eyer}}]{2019A&A...625A..97R}
{Rimoldini}, L., {Holl}, B., {Audard}, M., {et~al.} 2019, \aap, 625, A97,
  \dodoi{10.1051/0004-6361/201834616}

\bibitem[{{S{\'a}nchez} {et~al.}(2019){S{\'a}nchez}, {Dom{\'\i}nguez R.},
  {Lares}, {Beroiz}, {Cabral}, {Gurovich}, {Qui{\~n}ones}, {Artola}, {Colazo},
  {Schneiter}, {Girardini}, {Tornatore}, {Nilo Castell{\'o}n}, {Garc{\'\i}a
  Lambas}, \& {D{\'\i}az}}]{2019A&C....2800284S}
{S{\'a}nchez}, B., {Dom{\'\i}nguez R.}, M.~J., {Lares}, M., {et~al.} 2019,
  Astronomy and Computing, 28, 100284, \dodoi{10.1016/j.ascom.2019.05.002}

\bibitem[{{Sesar} {et~al.}(2010){Sesar}, {Ivezi{\'c}}, {Grammer}, {Morgan},
  {Becker}, {Juri{\'c}}, {De Lee}, {Annis}, {Beers}, {Fan}, {Lupton}, {Gunn},
  {Knapp}, {Jiang}, {Jester}, {Johnston}, \& {Lampeitl}}]{2010ApJ...708..717S}
{Sesar}, B., {Ivezi{\'c}}, {\v{Z}}., {Grammer}, S.~H., {et~al.} 2010, \apj,
  708, 717, \dodoi{10.1088/0004-637X/708/1/717}

\bibitem[{{Smolec}(2005)}]{2005AcA....55...59S}
{Smolec}, R. 2005, \actaa, 55, 59.
\newblock \doarXiv{astro-ph/0503614}

\bibitem[{{Spergel} {et~al.}(2015){Spergel}, {Gehrels}, {Baltay}, {Bennett},
  {Breckinridge}, {Donahue}, {Dressler}, {Gaudi}, {Greene}, {Guyon}, {Hirata},
  {Kalirai}, {Kasdin}, {Macintosh}, {Moos}, {Perlmutter}, {Postman},
  {Rauscher}, {Rhodes}, {Wang}, {Weinberg}, {Benford}, {Hudson}, {Jeong},
  {Mellier}, {Traub}, {Yamada}, {Capak}, {Colbert}, {Masters}, {Penny},
  {Savransky}, {Stern}, {Zimmerman}, {Barry}, {Bartusek}, {Carpenter}, {Cheng},
  {Content}, {Dekens}, {Demers}, {Grady}, {Jackson}, {Kuan}, {Kruk}, {Melton},
  {Nemati}, {Parvin}, {Poberezhskiy}, {Peddie}, {Ruffa}, {Wallace}, {Whipple},
  {Wollack}, \& {Zhao}}]{2015arXiv150303757S}
{Spergel}, D., {Gehrels}, N., {Baltay}, C., {et~al.} 2015, arXiv e-prints,
  arXiv:1503.03757.
\newblock \doarXiv{1503.03757}

\bibitem[{{The PLAsTiCC team} {et~al.}(2018){The PLAsTiCC team}, {Allam},
  {Bahmanyar}, {Biswas}, {Dai}, {Galbany}, {Hlo{\v{z}}ek}, {Ishida}, {Jha},
  {Jones}, {Kessler}, {Lochner}, {Mahabal}, {Malz}, {Mand el},
  {Mart{\'\i}nez-Galarza}, {McEwen}, {Muthukrishna}, {Narayan}, {Peiris},
  {Peters}, {Ponder}, {Setzer}, {The LSST Dark Energy Science Collaboration},
  {LSST Transients}, \& {Variable Stars Science
  Collaboration}}]{2018arXiv181000001T}
{The PLAsTiCC team}, {Allam}, Tarek, J., {Bahmanyar}, A., {et~al.} 2018, arXiv
  e-prints, arXiv:1810.00001.
\newblock \doarXiv{1810.00001}

\bibitem[{{Tr{\"o}ster} {et~al.}(2019){Tr{\"o}ster}, {Ferguson},
  {Harnois-D{\'e}raps}, \& {McCarthy}}]{2019MNRAS.487L..24T}
{Tr{\"o}ster}, T., {Ferguson}, C., {Harnois-D{\'e}raps}, J., \& {McCarthy},
  I.~G. 2019, \mnras, 487, L24, \dodoi{10.1093/mnrasl/slz075}

\bibitem[{{Tsang} \& {Schultz}(2019)}]{2019ApJ...877L..14T}
{Tsang}, B. T.~H., \& {Schultz}, W.~C. 2019, \apjl, 877, L14,
  \dodoi{10.3847/2041-8213/ab212c}

\bibitem[{{Udalski} {et~al.}(1992){Udalski}, {Szymanski}, {Kaluzny}, {Kubiak},
  \& {Mateo}}]{1992AcA....42..253U}
{Udalski}, A., {Szymanski}, M., {Kaluzny}, J., {Kubiak}, M., \& {Mateo}, M.
  1992, \actaa, 42, 253

\bibitem[{{Udalski} {et~al.}(2008){Udalski}, {Szymanski}, {Soszynski}, \&
  {Poleski}}]{2008AcA....58...69U}
{Udalski}, A., {Szymanski}, M.~K., {Soszynski}, I., \& {Poleski}, R. 2008,
  \actaa, 58, 69.
\newblock \doarXiv{0807.3884}

\bibitem[{{van der Walt} {et~al.}(2011){van der Walt}, {Colbert}, \&
  {Varoquaux}}]{2011CSE....13b..22V}
{van der Walt}, S., {Colbert}, S.~C., \& {Varoquaux}, G. 2011, Computing in
  Science and Engineering, 13, 22, \dodoi{10.1109/MCSE.2011.37}

\bibitem[{{W}es {M}c{K}inney(2010)}]{mckinney-proc-scipy-2010}
{W}es {M}c{K}inney. 2010, in {P}roceedings of the 9th {P}ython in {S}cience
  {C}onference, ed. {S}t\'efan van~der {W}alt \& {J}arrod {M}illman, 56 -- 61,
  \dodoi{10.25080/Majora-92bf1922-00a}

\bibitem[{{Yi} {et~al.}(2020){Yi}, {Guo}, {Fan}, {Hamann}, \&
  {Wang}}]{2020arXiv200111651Y}
{Yi}, K., {Guo}, Y., {Fan}, Y., {Hamann}, J., \& {Wang}, Y.~G. 2020, arXiv
  e-prints, arXiv:2001.11651.
\newblock \doarXiv{2001.11651}

\bibitem[{{Yu} \& {Koltun}(2015)}]{2015arXiv151107122Y}
{Yu}, F., \& {Koltun}, V. 2015, arXiv e-prints, arXiv:1511.07122.
\newblock \doarXiv{1511.07122}

\bibitem[{{Zhang}(2019)}]{2019arXiv190304687Z}
{Zhang}, L. 2019, arXiv e-prints, arXiv:1903.04687.
\newblock \doarXiv{1903.04687}

\bibitem[{{Zorich} {et~al.}(2020){Zorich}, {Pichara}, \&
  {Protopapas}}]{2020MNRAS.492.2897Z}
{Zorich}, L., {Pichara}, K., \& {Protopapas}, P. 2020, \mnras, 492, 2897,
  \dodoi{10.1093/mnras/stz3426}

\end{thebibliography}

\end{document}